\documentclass[twocolumn,prx,apssuperscriptaddress,nofootinbib]{revtex4}
\usepackage{graphicx}
\usepackage{bbold}
\usepackage{amsmath,amssymb}
\usepackage{float}
\usepackage{color}
\usepackage{enumitem}

\usepackage[utf8]{inputenc}
\usepackage[T1]{fontenc}
\usepackage{mathptmx}

\def\be{\begin{eqnarray}}
\def\ee{\end{eqnarray}}
\def\ben{\begin{eqnarray*}}
\def\een{\end{eqnarray*}}
\def\bes{\begin{subequations}}
\def\ees{\end{subequations}}
\def\ds{\displaystyle}
\def\nn{\nonumber}

\def\bes{\begin{subequations}}
  \def\ees{\end{subequations}}

\newcommand{\wig}[1]{\mathrel{\hbox{\hbox to 0pt{\lower.6ex\hbox{$\sim$}\hss    }\raise.4ex\hbox{$#1$}}}}

\begin{document}

\title{Universal Character of Atomic Motions at the Liquid-Solid Transition}

\author{J\'er\^ome \surname{Daligault}}
\email{daligaul@lanl.gov}
\affiliation{Los Alamos National Laboratory, Los Alamos, NM 87545, USA}

\begin{abstract}
We show evidence from computer simulations of a universal feature in
the atomic dynamics of simple liquids that heralds the freezing
transition.  This finding provides new insights into what changes at
the atomic level as the freezing point is traversed and allows the
system to discover the crystalline order.  We find that the
first-passage properties of atoms at the freezing point, namely the
mean time ${\cal{T}}(r)$ for an atom to first reach a distance $r$
from its initial position and the associated probability
distributions, are insensitive to the nature of the interparticle
force law.  For temperatures above freezing, the mean first-passage
time ${\cal{T}}(r)$ behaves as $r^{D(r)}$ with a power index $D(r)$
that monotonically increases from $D(r)\!=\!1$ at small $r$
(free-particle behavior) to $D(r)\!=\!2$ at large $r$ (diffusive
behavior).  At freezing, and regardless the nature of interactions,
$D(r)$ no longer varies monotonically between these two values but
exhibits a peak of height $D(r_*)\!=\!2.1$ at some distance $r_*$.
This behavior suggests that the location of the freezing transition is
concomitant with a universal degree of localization of atomic motions
above which the delicate balance between the disordering effects of
thermal agitation and the ordering effects of interactions can be
destabilized in favor of the periodic order.  To help understand and
quantitatively characterize the underlying physics, we develop a model
of the first-passage properties of atomic motions in liquids.  The
model builds on the potential energy landscape theory for liquids
according to which the liquid’s configuration vibrates for a time
around a stable local minimum on the potential energy surface and
occasionally transits to an adjacent minimum on the surface.  The
model faithfully reproduces the key features of the first passage time
properties observed in the computer simulations.  The model implies
that, at the freezing temperature, the average time $\tau$ separating
two transits on the potential energy surface equals the average period
of oscillation $\tau_o$ of atoms in the local minimum; for
temperatures above (below) freezing, $\tau$ is smaller (larger) than
$\tau_o$.  As a practical consequence of this work, we demonstrate
that the calculation of $D(r)$ gives rise to an efficient method for
determining the liquid-solid coexistence curves of real materials from
atomistic simulations, which, unlike other methods, does not require
knowing the crystalline structure of the solid phase.
\end{abstract}

\date{\today}

\maketitle

\section{Introduction} \label{section_1}

Nearly all fluids freeze into a periodic structure when gently cooled
or compressed.  The conventional view of the freezing transition holds
that, passed a particular point, the liquid state becomes metastable
with respect to the crystalline state and small crystal embryos
spontaneously form and re-dissolve via stochastic thermal
fluctuations, unless their size exceeds a critical value beyond which
they irreversibly grow and coalesce \cite{1,7}.  Yet, we understand
little about the microscopic processes underlying these spontaneous,
stochastic events.  Pure liquids can generally be super-cooled or
over-compressed passed these conditions with no sign of abrupt changes
in their properties around the transition \cite{3,5}.  For weak
undercooling, the time required to form a stable solid nucleus is
often too long to capture experimentally or in computational studies
unless an external disturbance substitute for spontaneous fluctuations
to initiate the transition.  Little is known about what actually
changes at the microscopic level at the freezing point that allows the
system to discover the crystalline order.  Evidence for distinguishing
structural and dynamical features in weakly under-cooled fluids is
tenuous and mainly limited to hard-sphere or Lennard-Jones fluids
\cite{4,6,Giaquinta1992} and colloidal liquids \cite{Lowenetal1993},
as is the evidence that such features are precursory to the formation
of stable crystal.  Nevertheless, the well-established Hansen-Verlet
freezing criterion \cite{9,10}, which states that a monoatomic liquid
freezes when the main peak of its structure factor $S(k)$ reaches a
universal value, namely $S(k)_{max}\simeq 2.85$, shows there are
aspects of the freezing transition that go beyond the details of
microscopic interactions.

In this work, we show evidence from computer simulations of a universal feature in the single particle dynamics of monatomic liquids that heralds the freezing transition.
This dynamical signature is related to the first passage properties of atomic motions, i.e. to the statistical distribution of the time taken by atoms to move by a certain distance from their initial position.
The numerical simulations suggest that the location of the freezing transition is concomitant with a common degree of localization of atomic motions.
In order to qualitatively characterize the latter, we develop a model of the first-passage properties (also referred to as first exit properties) that builds on the accepted picture according to which a liquid’s configuration vibrates for a time about a local minimum of the many-body potential energy surface and occasionally transits to an adjacent minimum on the surface. 
We find that the universal feature corresponds to conditions where the average time $\tau$ separating two transits is equal to the average period of oscillation $\tau_o$ of an atom about an equilibrium position.
When entering the undercooled regime, $\tau>\tau_0$ and atoms remain localized for times longer than the typical period of oscillations in the local potential energy valley.
We speculate that the longer localization in the valleys of the potential energy surface is a necessary condition for atoms to interact constructively and find the route to a local crystalline order.
In addition, we demonstrate that the precursory feature gives rise to a new practical method for determining the liquid-solid coexistence line of real materials from atomistic simulations.

\begin{figure}[t]
\begin{center}
\includegraphics[scale=0.4]{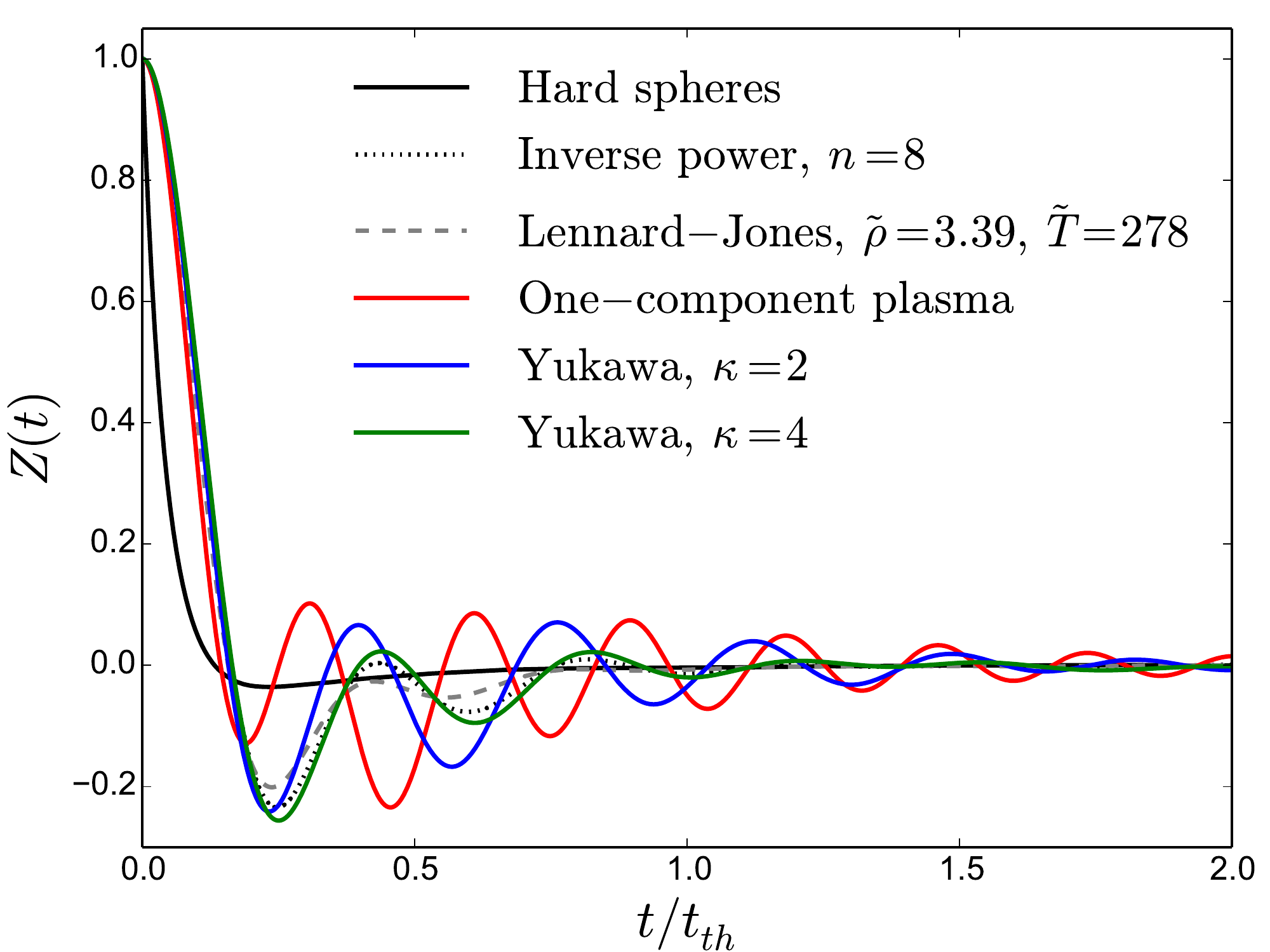}
\end{center}
\caption{(color online) Normalized velocity autocorrelation functions of reference models of liquids at their freezing conditions (see appendix~\ref{appendix_1}), including: the hard-sphere model, $v(r)=\infty$ for $r < \sigma$  and $0$ otherwise, where $\sigma$  is the particle diameter; the inverse power law or soft sphere model, $v(r)=\epsilon (r/\sigma)^n$ , with $n$ controlling the stiffness of particles; the Lennard-Jones model, $v(r)=4\epsilon[(r/\sigma)^{12}-(r/\sigma)^6 ]$, used to model fluids made of neutral atoms or small molecules; and the one-component plasma model, $v(r)=q^2 e^{-\kappa r}/r$, with $\kappa\geq 0$  controlling the range of $v(r)$, used to model ions in dense plasmas.
On the horizontal axis, the time is shown in units of $t_{th}=a/v_{th}$, where $a$ is the average interparticle distance and $v_{th}$ is the thermal velocity.
\label{figure_vafs}}
\end{figure}

\section{Signature of the freezing transition} \label{section_2}

Despite their structural similarity, the temporal dynamics of simple liquids at freezing does generally depend on the nature of interactions. 
This is illustrated in Fig.~\ref{figure_vafs} that shows the normalized velocity autocorrelation function (VAF) $Z(t)=\frac{m}{3k_B T}\left\langle\vec{V}_i(t)\cdot\vec{V}_i(0)\right\rangle_{eq}$ at the freezing point of several reference models of liquids characterized by distinct interaction potentials $v(r)$ (see figure caption).
For a fair comparison between the models, the time in Fig.~\ref{figure_vafs} is normalized to the reference time $t_{th}=a/v_{th}$ for a particle with thermal velocity $v_{th}=\sqrt{k_B T/m}$ to freely travel the mean interparticle distance $a=(3/4\pi\rho)^{1/3}$, where $T$ and $\rho$ are the temperature and the number density at freezing for each system (recalled in appendix \ref{appendix_1}).
In all cases, negative correlation regions develop caused by the localized oscillations of an atom in the cage formed by its immediate neighbors 
until the continuous dynamical rearrangement of particles leads to the disruption of the original shell of atoms and to the escape of the particle from its initial location  \cite{11}.
Although these models satisfy well the Hansen-Verlet criterion, the details of this local motions depend appreciably on $v(r)$. For instance, for hard spheres, $Z(t)$ rapidly vanishes after the first rebound against the initial cage, while for the Lennard-Jones interaction, $Z(t)$ oscillates with larger negative correlations than for hard-spheres. For the Coulomb ($\kappa=0$) one-component plasma, unlike other models, the lowest minimum of the VAF is attained by its second minimum. This is because, in addition to the oscillatory motions in the cages, particles also couple to the collective, high-frequency (plasma) charge oscillations \cite{11ocp}; this effect disappears with increasing $\kappa$ as the plasma oscillations are replaced by low-frequency sound waves.

\begin{figure}[t]
\begin{center}
\includegraphics[scale=0.4]{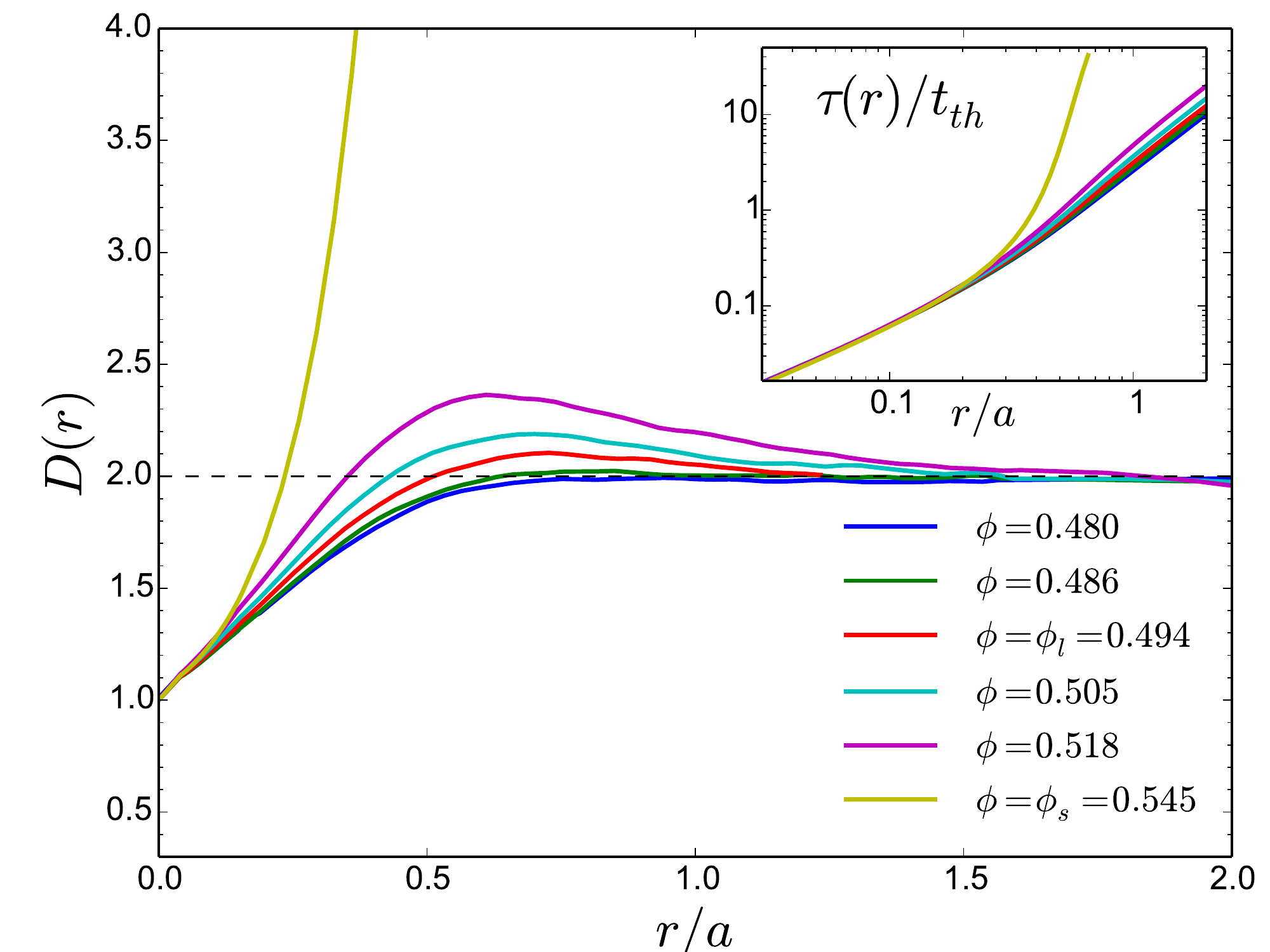}
\end{center}
\caption{(color online) Logarithmic derivative $D(r)=d\ln {\cal{T}}\!(r)/d\ln(r)$ of the mean first exit time for the hard-sphere system at various values of the packing fraction $\phi$. The inset shows the mean first exit time ${\cal{T}}\!(r)$.  In all cases, the system remains in the fluid phase as nucleation does not occur during the finite time of the simulations.
\label{figure_hs}}
\end{figure}
\begin{figure*}[t]
\begin{center}
\includegraphics[scale=0.4]{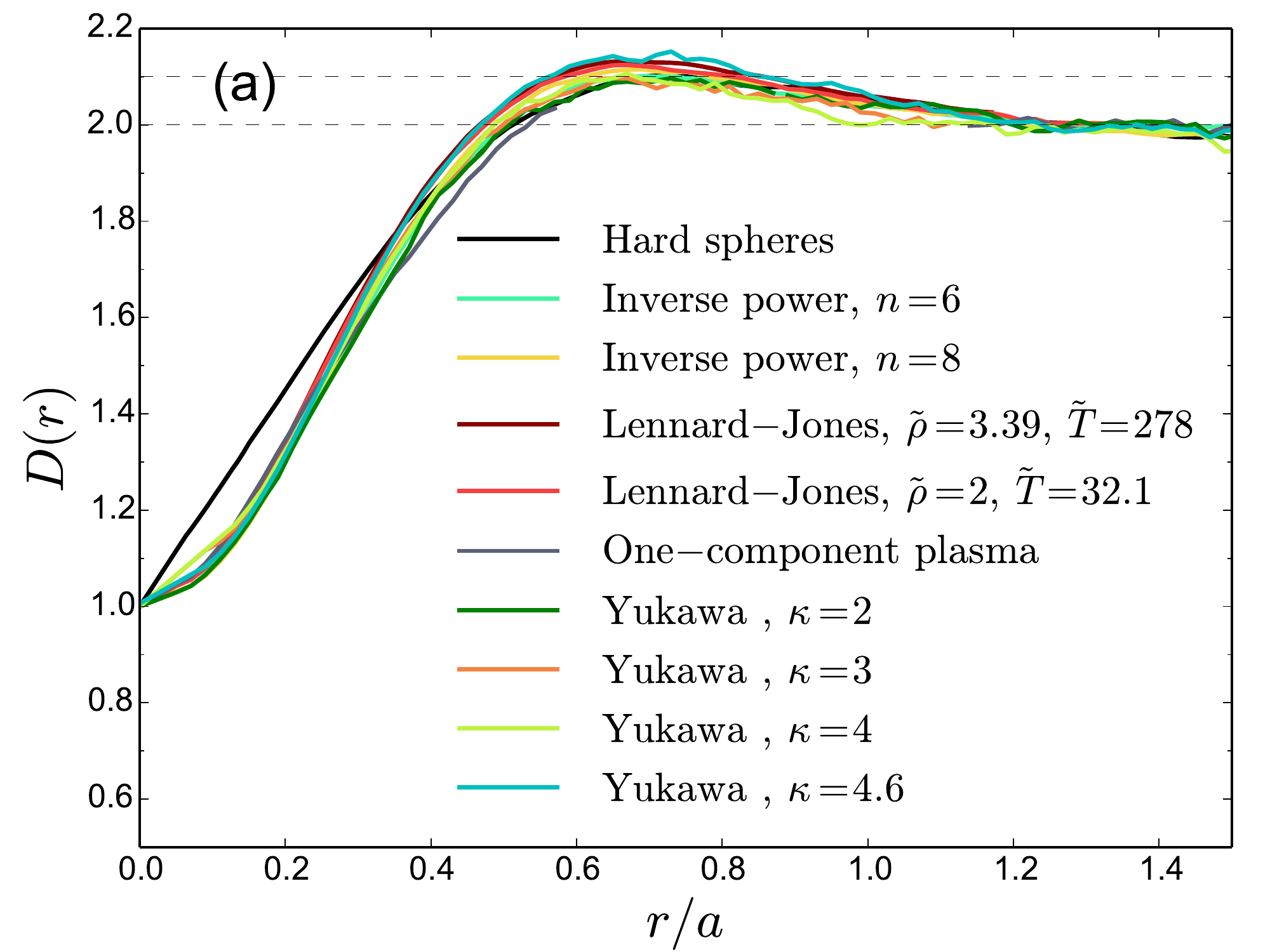}
\includegraphics[scale=0.4]{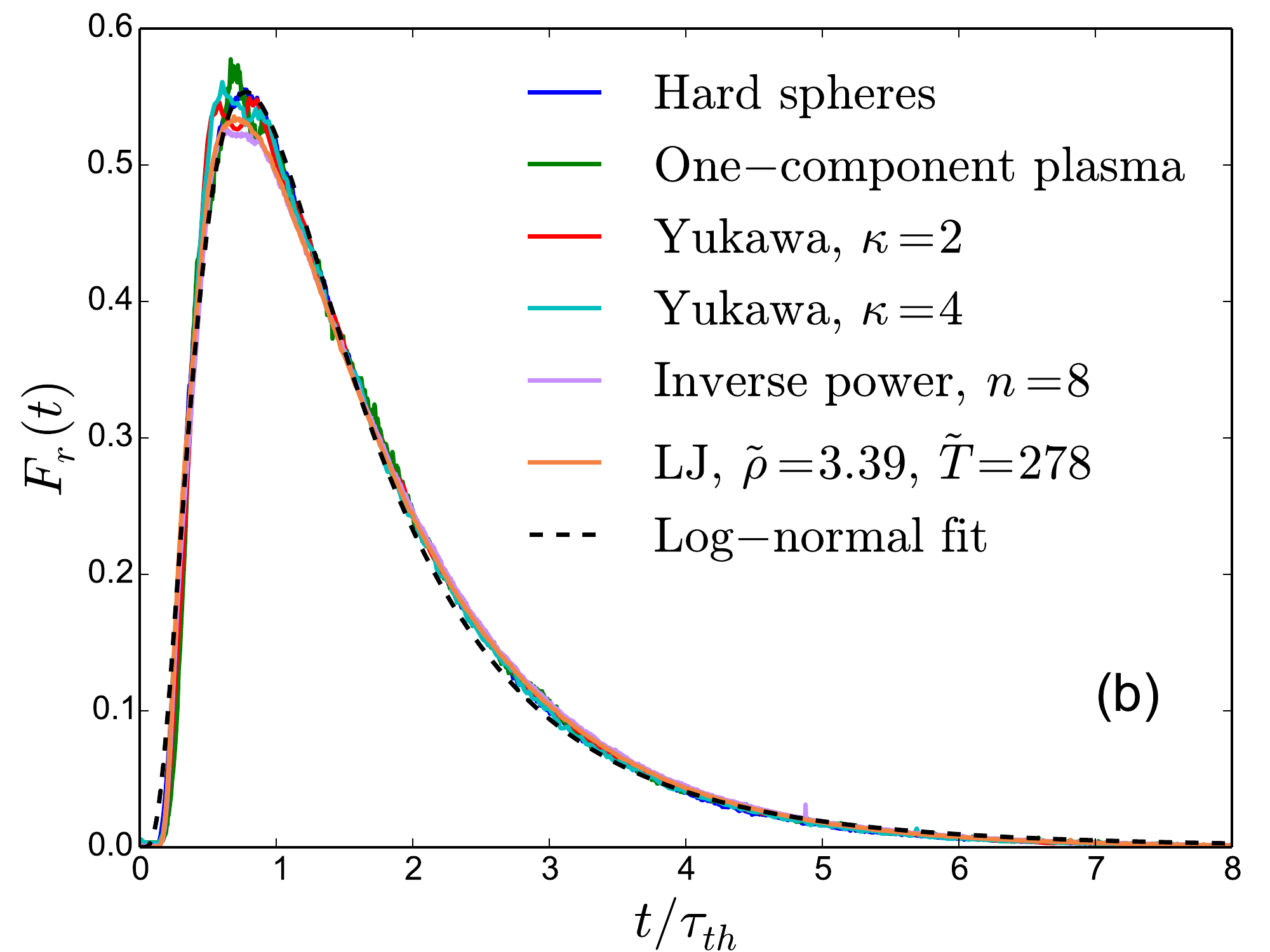}
\end{center}
\caption{(color online) (a) Logarithmic derivative $D(r)=d \ln {\cal{T}}\!(r)/d\ln(r)$  of the mean first exit time for several reference models of 3D liquids at their freezing point. The freezing points were determined by others from accurate free-energy calculations (see appendix~\ref{appendix_1}). (b) First-exit time probability distributions $f(r,t)$ for the 3D liquids of panel (a) measured at the location $r=r^*$ of the maximum of $D(r)$ (for clarity, only a subset of the cases in panel (a) are shown). The dashed line is a least-square fit to a log-normal distribution function.
\label{figure_allliquids}}
\end{figure*}
We will see that, despite the dependence of the VAF's on interparticle forces, their are aspects of the single particle dynamics that go beyond the details of interactions.
To this end, we will look at the dynamics from a different point of view.
In traditional studies on the dynamical properties of liquids, one typically follows the evolution of a dynamical variable $A(t)$ as a function of time $t$, e.g. the mean-square displacement, the VAF, etc. \cite{11,25}.
Here, we will instead consider the amount of time $t(A_*)$ required for the variable $A$ to first reach a threshold value $A_*$.
In particular, we will focus on the first-passage properties of atomic trajectories, i.e. on the statistical properties of the time it takes for an atom to first reach a distance $r$ from an initial position in the liquid \cite{13}.
More generally, the notion of first-passage times, or equivalently of first-hitting or first-exit times, plays an important role in many areas, including economics, mathematics, biology, physics \cite{12} and elsewhere, whenever a problem requires predicting the amount of time required for a stochastic process, e.g. the price of a stock option, starting from some initial state, to encounter a threshold for the first time.

We consider the probability distributions of first exit times $f(r,t)$ defined such that $f(r,t)dt$ is the probability that an atom reaches the distance $r$ from its initial position between times $t$ and $t+dt$. 
For illustration, the inset of Fig.~\ref{figure_hs} shows on a logarithmic plot the mean first exit time ${\cal{T}}(r)=\int_0^\infty{tf(r,t)dt}$  (in units of $t_{th}$) as a function of $r$ (in units of $a$) for the hard-sphere model at several packing fractions $0.48\le \phi\le 0.545$, with $\phi=\pi\rho\sigma^3/6$.
We recall that the phase diagram of hard spheres reduces to two vertical lines in the density-temperature plane: a fluid phase for packing fractions $\phi\le\phi_l=0.494$ and a solid phase for $\phi\ge \phi_s=0.545$, separated by a coexistence zone $\phi_l\le\phi\le\phi_s$. ${\cal{T}}(r)/t_{th}$ increases with $\phi$ since the displacements are more hampered by the more frequent collisions against neighbors. It also exhibits, in the language of exit times, the well-known transition in the particle displacements between the inertial motion regime at short-time scale and the diffusive motion regime at longer time, which imply ${\cal{T}}(r)\propto r$ for $r\rightarrow 0$ and as $r^2$ for large $r$. This transition is best seen in the variation shown in Figure~\ref{figure_hs} of the logarithmic derivative $D(r)=\frac{d\ln {\cal{T}}\!(r)}{d\ln(r)}$ , a dimensionless quantity that gives information on the power-law scaling behaviour of ${\cal{T}} (r)$ with $r$ (the diffusive limit is not fully reached over the range of distances shown here, especially at large $\phi$). Most importantly here, Figure~\ref{figure_hs} shows that $\phi=\phi_l$ separates two regimes of particle caging. In the stable phase $\phi<\phi_l$, $D(r)$ increases monotonically between the inertial and diffusive limits since the cages are rapidly disrupted and the particles can easily wander off. At $\phi=\phi_l$, $D(r)$ shows a small hump at $r=r^*\simeq 0.7a$ of height $D(r^* )\simeq 2.1$, slightly higher than in the diffusive limit. For $\phi>\phi_l$, the hump becomes more pronounced and $r^*$ decreases with increasing $\phi$. This indicates the tendency of the particle motions to become increasingly spatially localized for longer time periods in the initial cage \cite{13}.

We have calculated $f(r,t)$ for various simple liquids using classical molecular dynamics simulations (details on the calculations are given in appendix~\ref{appendix_1}).
Figure~\ref{figure_allliquids}a shows $D(r)$ at the freezing points of several reference models of liquids, including the cases used in Fig.~\ref{figure_vafs} to illustrate the dependence of the VAF's on the interaction potential.
The strong similarity between the plots in Fig.~\ref{figure_allliquids}a is striking.
In all cases, the height of the hump is within less than $2\%$ the hard-sphere value $2.1$.
In addition, whereas $D(r)$ relates to an average property of first-exit times, we find that the probability distributions $f(r,t)$  themselves are insensitive to the interatomic forces.
Figure~\ref{figure_allliquids}b shows $f(r,t)$  against the reduced time $t/t_{th}$  at the peak position $r=r^*$ of $D(r)$.
The agreement between different liquids is remarkable and says that the probability for any atom of the liquid to reach the `cage radius' $r^*$ in any time $t/t_{th}$ is nearly independent of the interaction potential.
\begin{figure}[t]
\begin{center}
\includegraphics[scale=0.5]{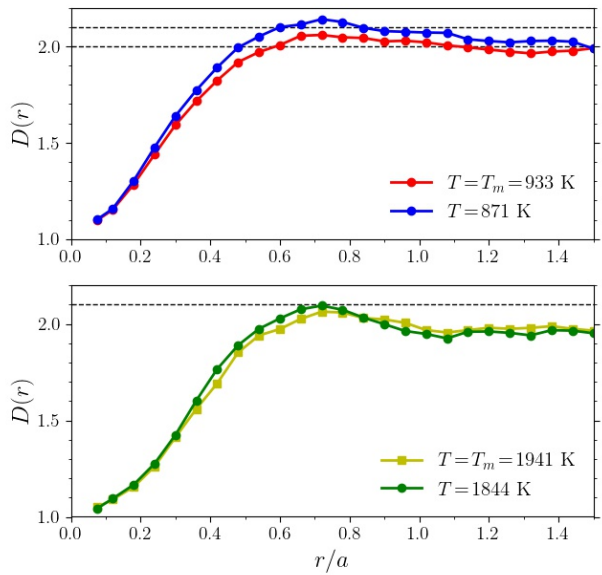}
\end{center}
\vspace*{-.5cm}
\caption{(color online) Logarithmic derivative $D(r)=d\ln(\tau)/d\ln(r)$ of the mean first exit time for liquid Aluminum (top panel) and liquid Titanium (bottom pane) at and slightly below their melting temperature $T_m$. Here the data were calculated from quantum molecular dynamics simulations.
  \label{fig_DofR_Al_qmd}}
\end{figure}
We also remark that these findings are insensitive to the symmetry of the stable crystal lattice selected: in Figure~\ref{figure_allliquids}a, some systems freeze into an FCC structure (e.g., hard spheres, Lennard-Jones, $\kappa =4.6$ Yukawa), others into a BCC structure (e.g. one-component plasma, Yukawa with $\kappa\le 4$). 
Because it is based on a measure of particle motions, the criterion can be regarded as the counterpart to freezing of the celebrated Lindemann criterion of melting \cite{15,10} that states that a crystal melts when the root mean-square displacement of atoms reaches a fraction $f=0.15$ of the nearest neighbor distance. We note that, in practice, $f$ varies more appreciably between systems (e.g., $f\simeq 0.133$ for hard spheres, $f\simeq 0.186$ for the one-component plasma \cite{16})  and depends on the crystal structure.

\begin{figure}[t]
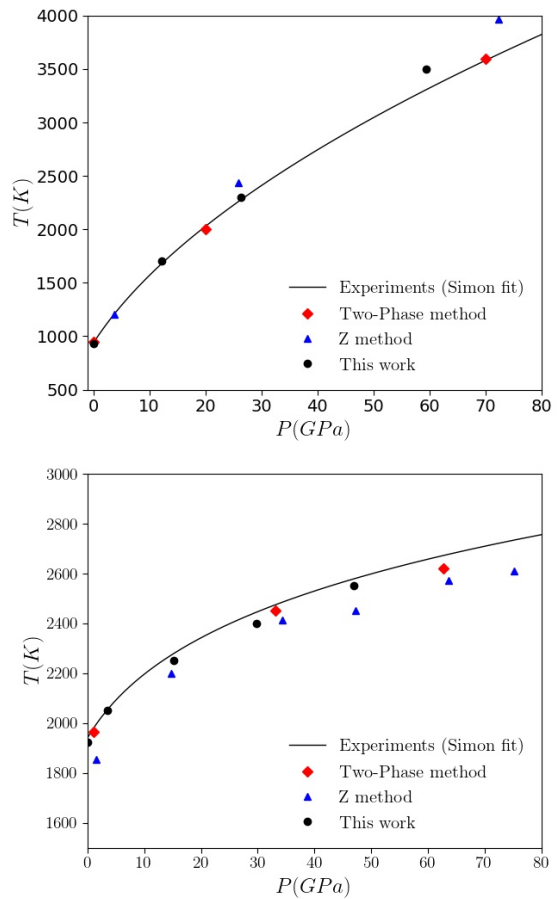

\begin{center}
  \includegraphics[scale=0.45]{PT_Al_QMD.pdf}\\
  \includegraphics[scale=0.45]{PT_Ti_QMD.pdf}
\end{center}
\caption{(color online) Liquid-solid coexistence line of Aluminum (top panel) and Titanium (bottom panel). All the symbols were obtained with quantum molecular dynamics simulations. The blacks dots show results of our newly proposed method based on characteristic behavior of $D(r)$ at the freezing transition. The red diamonds and blue diamonds show data reported in Refs.~\cite{Bouchet2009} and \cite{Stutzmann2015} and obtained using the coexistence method and the Z-method, respectively. The full lines show Simon fits of experimental measurements reported in Ref.\cite{Boehler1997} for Al (see their Fig.~1) and \cite{Stutzmann2015} for Ti (see their Fig.~4a).
\label{figure_liquid_solid_line}}
\end{figure}
So far, we have considered simple models of liquids where particles interact classically via a given potential $v(r)$ and move according to Newton's laws.
In order to test whether the characteristic behavior of $D(r)$ at freezing applies to real liquids, we have performed quantum molecular dynamics (QMD) simulations of elemental liquid metals.
In these simulations (see details appendix~\ref{appendix_1}), ions propagate classically in a periodic simulation cell, while electrons receive full quantum mechanical treatment using density functional theory and remain in the thermal ground state of the instantaneous ionic configurations (Born-Oppenheimer approximation).
Due to the higher computational cost of these simulations, a relatively small number of atoms $N=64$ was used, to be compared with $N=1000$ used in the classical simulations discussed previously.
Figure~\ref{fig_DofR_Al_qmd} shows $D(r)$ obtained for two different liquid metals, namely: Aluminum at melt density $2.35$ $\rm g.cm^{-3}$ for two temperature, the melting temperature $T=T_m=933$ K and $T=871$ K; and Titanium, a transition metal, at melt density $4.11$ $\rm g.cm^{-3}$ for two temperature, the melting temperature $T=T_m=1941$ K and $T=1844$ K.
In both cases, we find that the characteristic behavior of $D(r)$ at the freezing transition found in simple liquid models is also well satisfied by these two metals.
These results give us confidence into the physical significance of this finding.

By extrapolation, we suggest that the characteristic behavior can be used as a practical criterion for determining the liquid-solid coexistence curves (melting temperature $T$ vs pressure $P$) of real materials from atomistic simulations.
This illustrated in Fig.~\ref{figure_liquid_solid_line} (black dots).
The black dots show predicted points on the liquid-solid coexistence curve obtained for liquid Aluminum (top panel) and liquid Titanium (bottom panel) using density functional theory based QMD simulations.
The points were obtained as follows.
Given an input density, the pressure $P$ and the power index $D(r)$ were calculated for different temperatures 
The temperature shown in Fig.~\ref{figure_liquid_solid_line} correspond to those that reproduce the freezing criterion $D(r*)|_{\rm max}=2.1$.
We find that our approach gives results in quite good agreement with the experiments (full lines in the figure).
Moreover, it competes with two of the most standard methods used to determine the coexistence curves of materials, namely: the Z-method (blue triangle), which relies on the limit of superheating of the solid phase; and the coexistence method (red diamonds), in which one monitors the evolution of the liquid phase in contact with the liquid phase.
We note that, unlike other methods, our method does not require knowing the crystalline structure of the solid, which is often a challenge in itself.
Moreover, both the Z-method and the coexistence methods require larger systems.

\section{Understanding the first passage properties of liquids at freezing} \label{section_4}

The previous findings suggest a strong correlation between the location of the freezing transition and the onset of a regime of localization of atomic motions.
To help understand and quantitatively characterize the phenomenon, we have developed a model of the first-passage properties of atoms in liquids.
To this end, we first establish an exact relation (Eq.(\ref{general_eq_for_f_integrated}) below) between the desired probability density of first passages $f(r,t)$ to the probability density $G_s({\bf r},t|{\bf r}_0)$ that an atom is at position ${\bf r}$ at time $t$ if it was initially located at ${\bf r}_0$.
We then develop a model for $G_s$ presented in Sec.~\ref{section_4_random_walk}, which combines an accurate description of the localized oscillations of an atom about an equilibrium position together with a continuous time random walk to account for the occasional jumps that occur between equilibrium positions.
The implications of the resulting model are discussed in Sec.~\ref{section_4_1_applications}. 

\subsection{Path to the distribution of first passages} \label{section_4_1}

\begin{figure}[t]
\vspace{0.2cm}
\begin{center}
\includegraphics[scale=0.4]{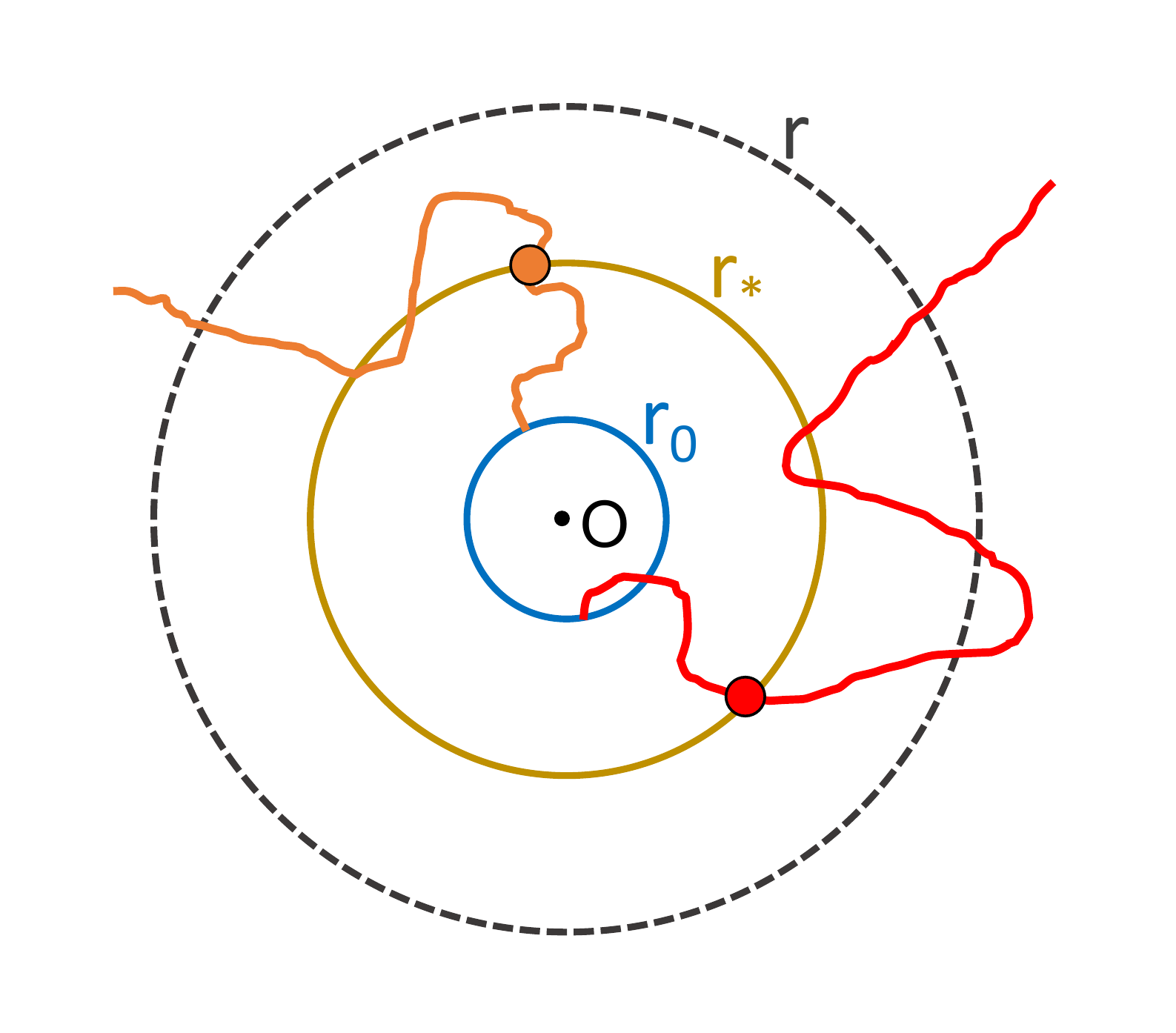}
\end{center}
\vspace{-0.4cm}
\caption{(color online) Distances involved in the definition of probability densities ${\cal{P}}(r,t|r_0)$ and ${\cal{P}}(r,t|r_*,t';0)$. The red and orange dots indicate the first passage at distance $r_*$ of an atom initially located on a shell of radius $r_0$ from some origin O.
\label{figure_model_cartoon_2}}
\end{figure}

Let us suppose for now that we know the following two quantities (see Fig.~\ref{figure_model_cartoon_2}):\\
\hspace*{0.5cm} 1) ${\cal{P}}(r,t|r_0)$ : the probability density that an atom initially located at a distance $r_0\!\geq\! 0$ from some origin O, be located at a distance $r\!\geq \! 0$ from O after a time $t$.\\
\hspace*{0.5cm} 2) ${\cal{P}}(r,t|r_*,t';0)$: the probability density that an atom is  located at a distance $r\geq 0$ from the origin O at time $t$ if it was both located at the origin O at time $0$ and at a distance $r_*\geq 0$ at some intermediate time $t'$ with $0\leq t'\leq t$.\\
For all distances $r$ and $r_*$ such that $r\geq r_*$, the two probability densities satisfy the relation
\be
{\cal{P}}(r,t|0)=\int_0^t{dt'{\cal{P}}(r,t|r_*,t';0)f(r_*,t')} \label{general_eq_for_f}
\ee
where $f(r_*,t')$ is the probability density for the first passage of an atom at a distance $r_*$ from its initial position, the quantity of interest in this work.
The relation (\ref{general_eq_for_f}) expresses that, the trajectories being continuous, for an atom to reach the distance $r$ at time $t$ from its initial position, it must necessarily be located at the distance $r_*\leq r$ at least once between the times $0$ and $t$; the time $t'$ corresponds to the time of its very first passage at $r_*$ (see red and orange dots in Fig.~\ref{figure_model_cartoon_2}).
Equation (\ref{general_eq_for_f}) is an integral equation for the unknown function $f$, a Volterra equation of the first kind \cite{PolyaninManzhirov_book}.
The probability of first passage $f(r_*,t)$ can in principle be found by solving Eq.(\ref{general_eq_for_f}) for any fixed distance $r\geq r_*$.
In practice, however, Eq.(\ref{general_eq_for_f}) is rather hard to solve because its kernel is singular, and we found it easier to instead consider the integral equation
\be
\int_{r_*}^\infty{dr\,{\cal{P}}(r,t|0)}=\int_0^t{dt'\left[\int_{r_*}^\infty{dr\,{\cal{P}}(r,t|r_*,t';0)}\right]f(r_*,t')}\nn\\
\label{general_eq_for_f_integrated}
\ee
obtained by summing Eq.(\ref{general_eq_for_f}) for all $r\geq r_*$.
The Volterra equation (\ref{general_eq_for_f_integrated}) has a smooth kernel and, unlike Eq.(\ref{general_eq_for_f}), lends itself to standard analytical and numerical methods \cite{PolyaninManzhirov_book}.

We now discuss in more details the two input probability densities.
For atoms in a liquid, ${\cal{P}}(r,t|0)\!=\!4\pi r^2 G_s(r,t)$ where $\ds G_s(r,t)\!=\!\left\langle\delta\left({\bf r}-[{\bf r}_i(t)-{\bf  r}_i(0)]\right)\right\rangle_{\rm eq}$ is the so-called self-correlation function (also know as the self-part of the Van Hove function) that is often used to characterize the single particle dynamics in liquids.
In the last expression, ${\bf r}_i(t)$ denotes the position of an atom $i$ at time $t$ and the brackets indicate an ensemble average.
For our purpose, it is useful to define the related quantity $G_s({\bf r},t|{\bf r}^\prime)=V\left\langle\delta\left({\bf r}-{\bf r}_i(t)\right)\delta\left({\bf r}^\prime-{\bf r}_i(0)\right)\right\rangle_{eq}$, which corresponds to the probability density that, if an atom i is initially at position ${\bf r}'$, it will be found at position ${\bf r}$ at time $t$.
In a liquid in thermal equilibrium, translational invariance implies the relation $G_s({\bf r},t|{\bf r}^\prime)=G_s(||{\bf r}-{\bf r}'||,t)$.
With this definition, 
\be
{\cal{P}}(r,t|r_0)=4\pi r^2 \int{d{\bf r}'\, G_s({\bf r},t|{\bf r}')p_{r_0}({\bf r}')} \label{calP_Gs}
\ee
where $p_{r_0}({\bf r})=\frac{1}{4\pi r_0^2}\delta(r-r_0)$ is the uniform probability density for an atom to be initially on a shell of radius $r_0$ from the origin O.

The probability density ${\cal{P}}(r,t|r_*,t';0)$ is harder to model than ${\cal{P}}(r,t|0)$ because the dependence on the two times $t'$ and $0$ requires a finer knowledge of the correlations governing the underlying particle dynamics.
Here, we shall make the markovian approximation that ${\cal{P}}(r,t|r_*,t';0)$ depends on the history only through the largest time $t'$ and not on the initial time, which gives
\be
{\cal{P}}(r,t|r_*,t';0)={\cal{P}}(r,t-t'|r_*)\,. \label{calPmarkovian}
\ee
This approximation relies on the effectiveness of many-particle interactions to destroy the dependence on the entire history (for non-interacting particles, the makovian approximation dramatically fails as the ballistic motion of an atom depends on its own initial position and velocity only).
The markovian approximation (\ref{calPmarkovian}) is expected to be increasingly accurate with increasing separation $r-r_*$, which is governed by the diffusive motion of particles.
Since, in addition, the kernel in Eq.(\ref{general_eq_for_f_integrated}) involves an integral over all $r\geq r_*$, we expect the kernel to be dominated by values of $r$ where the markovian approximation is good.
This is favorably tested below by comparing the results of the model against molecular dynamics results.

With Eq.(\ref{calPmarkovian}), the Volterra equation (\ref{general_eq_for_f_integrated}) for $f$ becomes
\be
g(r,t)=\int_0^t{dt' K(r,t-t')f(r,t')} \label{markovian_eq_for_f_integral}
\ee
where $g(r,t)=\int_{r}^\infty{dr'\,{\cal{P}}(r',t|0)}$ and the kernel $K(r,t-t')=\int_{r}^\infty{dr'\,{\cal{P}}(r,t-t'|r')}$ depends on the time difference $t-t'$.
The Volterra equation (\ref{markovian_eq_for_f_integral}) can then be formally be solved using Laplace transforms to give
\be
\hat f(r,s)=\frac{\int_{r}^\infty{dr\,\hat{\cal{P}}(r,s|0)}}{\int_{r}^\infty{dr\,\hat{\cal{P}}(r,s|r)}}=\frac{\hat g(r,s)}{\hat K(r,s)}\,. \label{markovian_Laplace_f}
\ee
One verifies that, despite the Markovian approximation (\ref{calPmarkovian}), the probability distribution $f$ defined by Eq.(\ref{markovian_Laplace_f}) is well normalized, i.e $\int_0^\infty{dt\/f(r,t)}=\lim_{s\to 0}\hat f(r,s)=1$.
Moreover, the mean first passage time of interest in this work is given by
\be
{\cal{T}}(r)=\int_0^\infty{dt\, tf(r,t)}=-\frac{\partial \hat f(r,s)}{\partial s}\Big|_{s=0}\,. \label{tau_R}
\ee
In the next section, we will present a model for $G_s({\bf r},t|{\bf r}_0)$ appropriate for physical conditions surrounding the liquid-solid transition.
We will then substitute this model into Eq.(\ref{markovian_eq_for_f_integral}) to extract information on $f(r,t)$ and on ${\cal{T}}(r)$.
Before, we conclude this section with an illustration of the approach on two simpler, yet instructive, models.

First, we assume that the atoms behave like Brownian particles in three dimensions.
In this case, $G_s$ satisfies the diffusion equation $\partial_tG_s=D\mathbf{\nabla}^2G_s$ with initial condition $G_s({\bf r},t=0|{\bf r}_0)=\delta({\bf r}-{\bf r}_0)$, where $D$ is the self diffusion coefficient, and reads
\be
G_s({\bf r},t|{\bf r}_0)=\frac{1}{(4\pi Dt)^{3/2}}e^{-({\bf r}-{\bf r}_0)^2/4Dt}\,. \label{Gs_diffusion}
\ee
Using Eq.(\ref{Gs_diffusion}) in Eq.(\ref{markovian_eq_for_f_integral}) \cite{details_on_diffusion_example}, we recover the known results for the three-dimensional Brownian motion obtained with other methods \cite{Klein1952,BorodinSalminenbook}, namely
\ben
\hat f(r,s)=\frac {r\sqrt{s/D}}{{\rm sh}\left(r\sqrt{s/D}\right)}\,,
\een
and the mean first passage time (\ref{tau_R}) is
\ben
{\cal{T}}(r)=\frac{r^2}{6D}\,.
\een
Thus, the logarithmic derivative $D(r)=d\ln {\cal{T}}(r)/d\ln r$ is constant equal to $2$ for all $r$, which is unlike the variation of $D(r)$ between $1$ and $2$ as $r$ increases observed previously for liquids (see Sec.~\ref{section_2}).

\begin{figure}[t]
\includegraphics[scale=0.5]{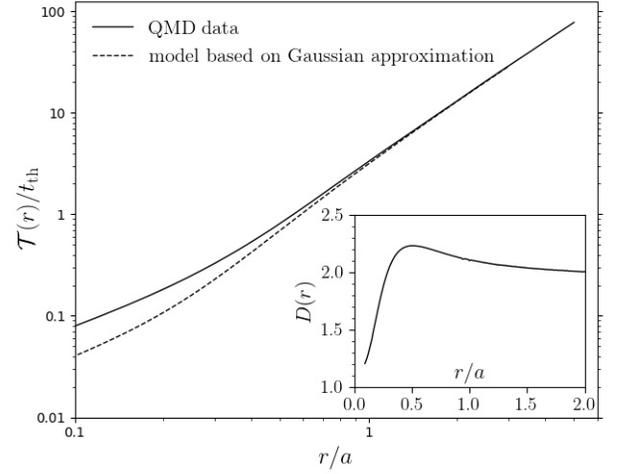}\\
\includegraphics[scale=0.5]{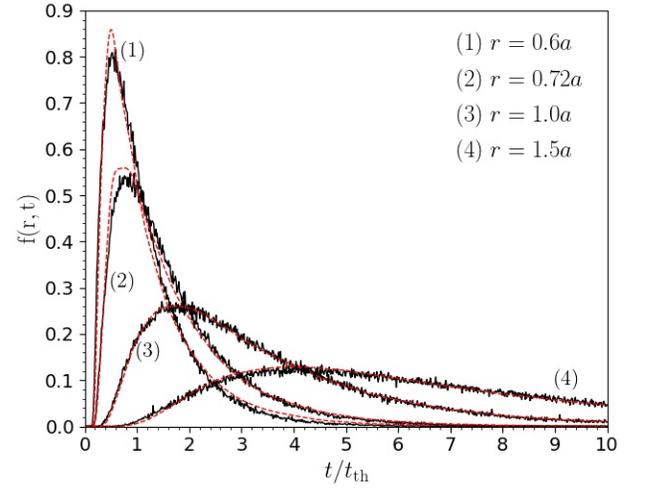}\\
\caption{(color online) The figure compares the first-passage properties of liquid Aluminum at $2.35$ $\rm g.cm^{-3}$ and $T=933$ K obtained with a QMD simulation (full lines) and from the numerical solution of Eq.(\ref{main_model_equation}) (dashed lines) with the width $W(t)$ calculated in the QMD simulation. Equation (\ref{main_model_equation}) was solved numerically using the trapezoidal rule \cite{BorodinSalminenbook}. Panel (a): first exit times ${\cal{T}}(r)$. The inset shows the power index $D(r)$ obtained with the model calculation.  Panel (b): first passage time probability distribution $f(r,t)$ vs $t$ for four values of $r$. For the QMD data, $r/a=0.6, 0.72, 1$ and $1.5$; for the model calculation, as explained in the text, the latter are shifted by $0.06$, i.e. $r/a=0.66, 0.78, 1.06$ and $1.56$.
\label{Fig_simple_model_Al}}
\end{figure}
As a second illustration, we consider a model that is more appropriate for describing the single particle dynamics of liquids.
It is based on the old observation that the van Hove self correlation function $G_s(r,t)$ for liquids is rather well approximated at all times by a Gaussian, namely
\be
G_s({\bf r},t|{\bf r}_0)=\left(\frac{3}{2\pi \langle R^2(t)\rangle}\right)^{\frac{3}{2}}\exp\left[-\frac{3}{2\langle R^2(t)\rangle}({\bf r}-{\bf r}_0)^2\right]\nn\\ \label{Gs_Gaussian}
\ee
where $\langle R^2(t)\rangle$ is the mean-square displacement.
This Gaussian approximation, which has been tested against MD data on various liquid models (e.g., \cite{Rahman1964,LevesqueVerlet1970,NijboerRahman1966}),  respects the exact limiting Gaussian behavior of $G_s$ at small times (free-particle behavior) and at large times (diffusive limit).
Using Eq.(\ref{Gs_diffusion}) in Eq.(\ref{markovian_eq_for_f_integral}), the integral equation for the first passage probability $f$ reads
\be
g(r_*,t)=\int_0^t{dt' K(r_*,t-t')f(r_*,t')} \label{main_model_equation}
\ee
with $ R\geq r_*$
\ben
g(r_*,t)&=&\left[1-{\rm erf}\left(\frac{r_*}{W(t)}\right)\right]+\frac{2}{\sqrt{\pi}}\frac{r_*}{W(t)}e^{-\frac{r_*^2}{W^2(t)}}\\
K(r_*,t)&=&\frac{W(t)}{2\sqrt{\pi} r_*}\left[1-e^{-\frac{4r_*^2}{W^2(t)}}\right]+\left[1-\frac{1}{2}{\rm erf}\left(\frac{2r_*}{W(t)}\right)\right]
\een
Figure \ref{Fig_simple_model_Al} shows results obtained by solving Eq.(\ref{main_model_equation}) with the mean-square displacement $\langle R^2(t)\rangle$ obtained in the QMD simulation of liquid Aluminum at $2.35$ $\rm g.cm^{-3}$ and $T=933$ $\rm K$ discussed previously in relation to Fig.~\ref{fig_DofR_Al_qmd}.
This calculation thus relies on only two approximations, namely the Gaussian approximation (\ref{Gs_Gaussian}) and the markovian approximation (\ref{calPmarkovian}).
In Fig.~\ref{Fig_simple_model_Al} (top panel), the mean time ${\cal{T}}(r)$ (dashed line) is compared to that obtained in the QMD calculation (full line).
Both calculations are in very good agreement for distances $r$ greater than the average interparticle distance $a$. 
This is because for large enough $r$, one essentially probes the diffusive regime where both the Gaussian and markovian approximations are accurate.
The largest discrepancies between the two calculations occur at the smallest distances $r$, i.e. the particle dynamics underlying the model calculation is on average slower than the actual dynamics.
At  very small $r$, when one basically probes the free particle motion, the Gaussian approximation remains accurate but the markovian approximation fails.
At intermediate distances, where the Gaussian approximation is the least accurate \cite{Rahman1964,LevesqueVerlet1970,NijboerRahman1966} and the markovian approximation is expected to become increasingly accurate, we see that the relative error quickly diminishes for $r/a\geq 0.5$.
As shown in the inset of Fig.~\ref{Fig_simple_model_Al}, this simple model calculation reproduces the characteristic evolution of the logarithmic derivative $D(r)$ between 1 (small $r$) and 2 (large $r$).
As a consequence of the inaccuracies, however, $D(r)$ peaks at a higher value, namely $2.2$, than the QMD data.

Finally, the black lines in Fig.~\ref{Fig_simple_model_Al} (bottom panel) show the probability distributions of first exit times $f(r,t)$ obtained in the QMD simulation for $r/a=0.6 ,0.72 ,1$ and $1.5$.
The red dashed lines show the solution of the integral equation (\ref{main_model_equation}) for $f(r+0.06a,t)$.
The shift was chosen in order for the peaks of the QMD and model calculations to coincide and accounts for the slower dynamics predicted by the model.
Yet, despite this shift, it is remarkable that the model reproduces quite well the general shape of the exact probability distributions.

The model we develop in the next section aims at reducing the inaccuracies of the Gaussian approximation at intermediate distances where the signature peak of $D(r)$ is located.

\subsection{Mixed random walk model of the self correlation function $G_s({\bf r},t;{\bf r}_0)$} \label{section_4_random_walk}

Our model builds on the generally accepted picture that an atom in a liquid undergoes a succession of quasiharmonic oscillations for a time about an equilibrium position as in a solid, which itself occasionally jumps from one equilibrium position to another.
This picture was introduced long ago (e.g., \cite{Frenkelbook}) to explain the experimental fact that the specific heat and the volume of a solid changes little, while its self-diffusion coefficient changes greatly, when it melts to a liquid.
The original description was later put on a firmer basis in terms of the motion of particle configurations on the many-body potential energy landscape \cite{StillingerWeber1984,Stillingerbook,Wallacebook}.
For a period of time, the liquid’s configuration oscillates harmonically about a local minimum of the many-body potential energy surface. Occasionally, the liquid configuration will have enough kinetic energy to cross a saddle point on the potential energy surface and will jump to the cell surrounding a different local minimum that is responsible for the diffusion and fluid flow \cite{StillingerWeber1984,Stillingerbook,Wallacebook}.
Much effort has been devoted to developing these ideas into theories of liquid dynamics, particularly theories of thermodynamics/self-diffusion in liquids and supercooled liquids \cite{Sears1965,Damle1968,Zwanzig1983,Keyes1997, RabaniGezelterBerne1997,Wallace1997,ChisolmClementsWallace2001}.
For later reference, we quote the celebrated model of Zwanzig \cite{Zwanzig1983}, who postulated a model for the velocity autocorrelation function
\be
Z(t)=Z_v(t)e^{-t/\tau}\,, \label{Z_Zwanzig}
\ee
where
\be
Z_v(t)=\frac{k_BT}{m}\int_0^\infty{d\omega \rho(\omega)\cos(\omega t)} \label{Zv}
\ee
is the velocity correlation of the quasiharmomic motion, where $\rho(\omega)$ is the normalized density of normal mode frequencies, and the factor $e^{-t/\tau}$ is caused y the jumps, where the `hopping time' $\tau$ is characteristic of the time between jumps/the lifetime which characterizes the distribution $e^{-t/\tau}$ of residence times in the cells.
The two most prominent ways to determine $\tau_{zw}$ are (a)  to extract it from the imaginary frequency INM distribution, developed most notably by Keyes, and (b) to set $\tau_{ZW}^{-1}$ equal to the long-time decay rate of the ‘‘cage correlation function’’ of Rabani, Gezelter, and Berne.

\begin{figure}[t]
\vspace{0.2cm}
\begin{center}
\includegraphics[scale=0.5]{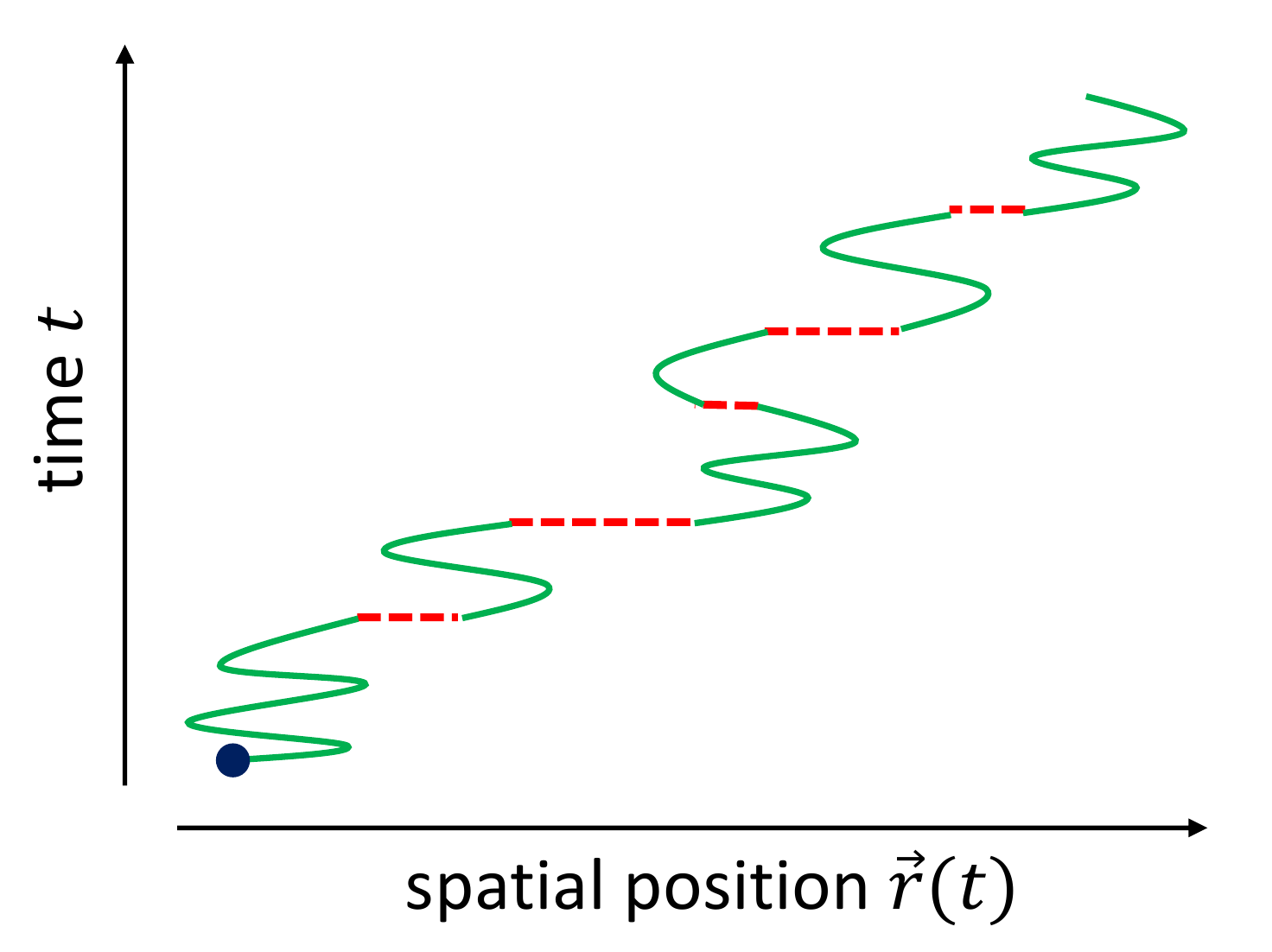}
\end{center}
\vspace{-0.4cm}
\caption{(color online) Schematic view of the random walk used in this work to model the self correlation (van Hove) function $G_s({\bf r},t;{\bf r}_0)$ of atoms in a liquid. The walk consists of quasiharmonic oscillations (green lines)  about an equilibrium position followed by occasional, instantaneous jumps (red dashes) from one equilibrium position to another at random times.
\label{figure3}}
\end{figure}
We develop a model for $G_s({\bf r},t|{\bf r}_0)$ that attempts to incorporate these observations with as few free parameters as possible.
The model combines a detailed description of the quasiharmonic oscillations about a local minimum of the potential energy surface together with a continuous time random walk (CRTW) to describe the finite size jumps between local potential energy minima on the surface.
The model is built from the following three components (see Fig.(\ref{figure3}):

1) {\it quasiharmonic motion}: we assume that the quasiharmonic oscillations about each local minimum is characterized by an average density of normal mode frequencies $\rho(\omega)$.
If one assumes that the atomic dynamics is limited to these oscillations only, a direct calculation shows that the self part of the van Hove function, which we denote by $f_v({\bf r},t)=\langle\delta({\bf r}-({\bf r}_i(t)-{\bf r}_i(0))\rangle$ (`v' stands for vibrations), remains Gaussian at all times and is given by
\be
f_v({\bf r},t)=\frac{1}{(\pi W_v^2(t))^{3/2}}e^{-r^2/W_v^2(t)} \label{f_v_Wv}
\ee
with the time dependent width 
\be
W_v^2(t)=\frac{4k_BT}{m}\int_0^\infty{d\omega\,\frac{\rho(\omega)}{\omega^2}\left[1-\cos(\omega t)\right]}\,. \label{Wv2_of_t}
\ee
The corresponding mean square displacement of an oscillating atom after time $t$ is $\langle \delta R^2(t)\rangle=\frac{3}{2}W_v^2(t)$.
Since the atoms do not diffuse, the latter reaches a constant value 
\be
r_v^2=\lim_{t\to\infty}\frac{3W_v^2(t)}{2}=\frac{6k_BT \langle \omega^{-2}\rangle}{m} \label{r_v}
\ee
at large times, where $\langle \omega^{-2}\rangle=\int_0^\infty{d\omega\omega^{-2}\rho(\omega)}$.
The distance $r_v$ can be regarded as the size of the cage in which an atom oscillates and will serve as a convenient unit of length in the following.
For later reference, we note the relation
\be
Z_v(t)=4\frac{d^2 W_v^2(t)}{dt^2}\,, \label{ZvandWv}
\ee
where the VAF $Z_v$ is given by Eq.(\ref{Zv}); this is a special case of the general relation
\be
Z(t)=\frac{1}{6}\frac{d^2}{dt^2}\langle \delta R^2(t)\rangle \label{general_relation_msd_vaf}
\ee
between the VAF and the mean square displacement \cite{25}. 

2) {\it Jumps between local minima}:
We assume the passage of the system from one valley of the potential energy surface to another occurs on a much shorter time scale than the typical oscillation time scale and that the system `jumps' instantaneously between equilibrium positions.
We assume that each such jump results in the displacement of some atoms from their current location.
We model these individual displacements by a continuous time random walk \cite{WeissRubin1983}, in which the size of the displacement is sampled from the Gaussian distribution
\ben
f_{J}({\bf r})=\frac{1}{(2\pi l^2)^{3/2}}e^{-r^2/2l^2}\,.
\een
of width $l$.

3) {\it Occurence of jumps}:
Finally, the term `continuous time' \cite{WeissRubin1983} indicates that the time interval between two successive jumps is also treated as a random variable.
To this end, we define $\phi(t)dt$ as the probability that the time interval separating two successive jumps is between $t$ and $t+dt$.
Then the quantity $\Psi(t) =1-\int_0^t{dt\phi(t)}$ is the probability that the time between two successive jumps is greater than $t$. 
Below \cite{note_on_fJ_phi}, we will assume that the jumps occur independently at a constant average rate $1/\tau$, i.e. they are distributed according to the exponential distribution
\be
\phi(t)=\frac{1}{\tau}e^{-t/\tau}\,. \label{phi_exponential}
\ee
We physically expect that $\tau$ increases with decreasing the system's temperature.

At this stage, the model contains three input parameters, namely $\rho(\omega)$, $l$ and $\tau$.
However, as we shall see later, in order for the model to be consistent with the dynamics of a liquid, $l$ is fact related to $\rho(\omega)$ and $\tau$, and there will only be two input parameters.

\begin{figure}[t]
\vspace{0.2cm}
\begin{center}
\includegraphics[scale=0.3]{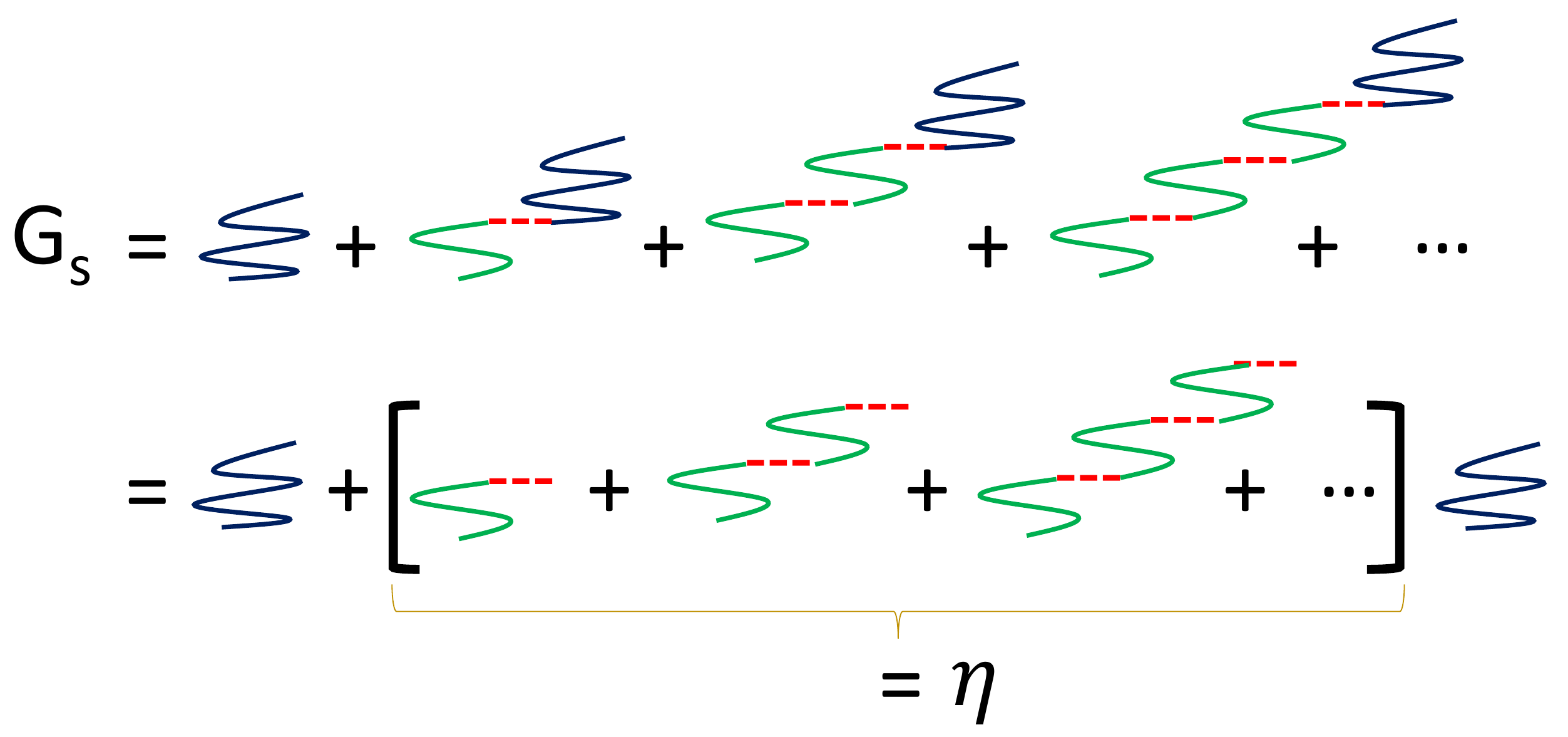}
\end{center}
\vspace{-0.4cm}
\caption{(color online) Illustration of the calculation of the probability density $G_s({\bf r},t;{\bf r}_0)$ that a particle is at position ${\bf r}$ at time $t\ge 0$ if it was at position ${\bf r}_0$ at time $t_0=0$. The contributions to $G_s$ are classified according to the number of quasiharmonic motions (green lines) followed by a jump (red dashes), and are then summed. The explicit expression of the second line is given by Eq.(\ref{G_s}), where $\eta$ is given by Eq.(\ref{eta}). Dropping variables and summation signs, the blue lines correspond to $f_v\Psi$ and the green and red lines correspond to $f_J\phi f_v$.
\label{figure_cartoon_3}}
\end{figure}
\begin{figure}[t]
\includegraphics[scale=0.45]{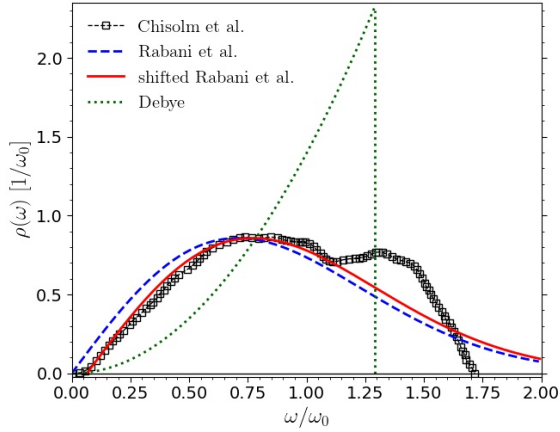}
\caption{(color online)
Symbols show the (dimensionless) density of normal mode frequencies $\rho(\omega)$ for liquid Na calculated with molecular dynamics simulations by Chisolm et al. (data taken from Fig.~1 in \cite{ChisolmClementsWallace2001}).
$\omega_0$ is the root mean squared frequency calculated with this density of states. The dashed blue line shows the model form $\rho(\omega)=2(\omega/\omega_0^2)e^{-\omega^2/\omega_0^2}$ proposed by  Rabani et al. \cite{RabaniGezelterBerne1997}. The red line shows the same curve but shifted horizontally to the right to remove the unphysical behavior at small frequencies of Rabani et al's model.
For reference, the dotted line shows the oversimplified Debye spectrum $\rho(\omega)\!=\!3\omega^2/\omega_D^3$ for $\omega\!<\!\omega_D\!=\!\sqrt{3/5}\,\omega_0$ and zero for $\omega\!>\!\omega_D$ used in early works for lack of a better choice (e.g., \cite{Zwanzig1983}).
\label{Fig:DofR_random_walk_Na}}
\end{figure}

\begin{figure}[t]
\includegraphics[scale=0.45]{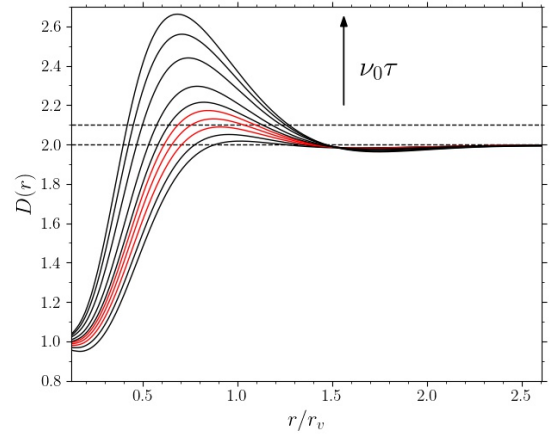}
\includegraphics[scale=0.45]{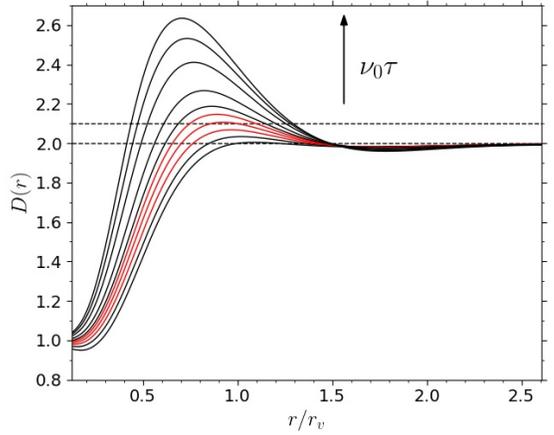}
\caption{(color online)
  Logarithmic derivatives $D(r)=d\ln{\cal{T}}(r)/d\ln(r)$ obtained by solving the model developed in Sec.~\ref{section_4} for the probability density of first passage $f(r,t)$ and the frequency spectra shown in Fig.~\ref{Fig:DofR_random_walk_Na}). Top panel: results obtained with spectrum $\rho(\omega)$ of Chisolm et al. (symbols in Fig.~\ref{Fig:DofR_random_walk_Na}). Bottom panel: results obtained with the modified spectrum of Rabani et al. (red line in Fig.~\ref{Fig:DofR_random_walk_Na}). Both panels show $D(r)$ for the same increasing values of the remaining parameter $\nu_0\tau=0.74,0.86,0.98,1.1,1.23,1.48,1.97,2.46$ and $2.96$. Curves highlighted in red correspond to $\nu_0\tau=0.86,0.98,1.11$.
\label{Fig:Dr_Na_rabani}}
\end{figure}

With these components defined, we wish to calculate the probability $G_s({\bf r},t|{\bf r}_0)$ that an atom is at position ${\bf r}$ at time $t\ge 0$ if it was at position ${\bf r}_0$ at time $t_0=0$.
There are infinitely many possible paths for an atom to go from ${\bf r}_0$ to ${\bf r}$, which can be distinguished by the number of jumps that occur in the time interval $[0:t]$ interspersed with local oscillatory motions.
One path involves {\it no} jumps, only the oscillatory motion that started since the last jump that occurred at $t=0$, and contributes $\Psi(t) f_v({\bf r}-{\bf r}_0,t)$ to $G_s({\bf r},t|{\bf r}_0)$.
There is then the paths that involve only one jump at time $t'$ in the time interval $[0\!:\!t]$, which contributes
\ben
\lefteqn{\iint{\!\!d{\bf r}_2d{\bf r}_1f_v({\bf r}-{\bf r}_2,t-t')\Psi(t-t')}}&&\\
&&\hspace{1.5cm}\times f_{J}({\bf r}_2-{\bf r}_1) \phi(t') f_v({\bf r}_1-{\bf r}_0,t')\,.
\een
Reading this term from the right side, the expression includes the oscillatory motion up to time $t'$ that takes the atom from ${\bf r}_0$ to ${\bf r}_1$ with probability $f_v({\bf r}_1-{\bf r}_0,t')d{\bf r}_1$, followed by a jump at $t'$ of length ${\bf r}_2-{\bf r}_1$ with probability $f_{J}({\bf r}_2-{\bf r}_1)d{\bf r}_2$, followed by the oscillatory motion between time $t'$ and $t$ that take the atom to postion ${\bf r}$.
The total contribution of all these single jump paths is obtained by integrating the previous term over $t'$ in $[0:t]$.
The classification of paths according to the number of intermediate jumps can be continued similarly and the total probability $G_s({\bf r},t|{\bf r}_0)$ is obtained by summing over all of them.
As illustrated in Fig.~\ref{figure_cartoon_3}, the sum of all possibilities resembles like a geometric sequence that can be summed into
\be
\lefteqn{G_s({\bf r},t|{\bf r}_0)=f_v({\bf r}-{\bf r}_0,t)\Psi(t)}&&\label{G_s}\\
&&\quad\quad+\int_0^t{dt'\int{d{\bf r}'f_v({\bf r}-{\bf r}',t-t')\Psi(t-t')\eta({\bf r}',t'|{\bf r}_0)}} \nn
\ee
where 
\be
\eta({\bf r},t|{\bf r}_0)&=&\int{d{\bf r}_1f_{J}({\bf r}-{\bf r}_1)\phi(t)f_v({\bf r}_1-{\bf r}_0,t)}\label{eta}\\
&+&\int_0^t{dt'\iint{d{\bf r}_1 d{\bf r}_2 f_{J}({\bf r}-{\bf r}_2) \phi(t-t')}}\nn\\
&&\hspace{2cm}\times f_v({\bf r}_2-{\bf r}_1,t-t')\eta({\bf r}_1,t'|{\bf r}_0)\nn
\ee
is the probability density that a jump was made to ${\bf r}$ between times $t$ and $t+dt$.
Equations (\ref{G_s}) and (\ref{eta}) imply that $G_s$ and $\eta$ are functions of the difference $|{\bf r}-{\bf r}_0|$, which permits the use of spatial Fourier transform with respect to the variable ${\bf r}-{\bf r}_0$.
With the exponential distribution (\ref{phi_exponential}), the Fourier-Laplace transform of Eq.(\ref{G_s}) takes the convenient compact form
\be
\hat G_s(k,s)=\frac{\ds \hat f_v\left(k,s+\frac{1}{\tau}\right)}{\ds 1-\frac{1}{\tau}\hat f_{J}(k) \hat f_v\left(k,s+\frac{1}{\tau}\right)}\,. \label{Gs_ks_model}
\ee
The prototypical Montroll-Weiss equation $\hat G_s=\tau/(\tau s+1-\hat{f}_J)$ \cite{WeissRubin1983} for a simple continuous-time random walk is recovered when removing the harmonic motions, $f_v({\bf r},t)=\delta({\bf r})$.
When the details of the oscillations are neglected, i.e. by replacing Eq.(\ref{f_v_Wv}) by a time-independent function $f_v({\bf r})=\frac{1}{(\pi w^2)^{3/2}}e^{-r^2/w^2}$, the model developed in Ref.~\cite{Chaudhuri2007} for supercooled liquids close to glass transition is recovered.

The expression (\ref{Gs_ks_model}) implies that, in this model, the mean square displacement (msd) $\left\langle\delta R^2(t)\right\rangle=\int{d{\bf r} r^2 G_s({\bf r},t;0)}$ is given by
\be
\left\langle\delta R^2(t)\right\rangle&=&\frac{3l^2}{\tau}t+\frac{3}{2}\int_0^t{du\int_0^u{dv \frac{d^2 W_v^2(v)}{dv^2}e^{-v/\tau}}} \label{msd_model}
\ee
The msd is asymptotic to $3l^2t/\tau$ as $t\to\infty$, i.e. the motion is diffusive at large times characterized by the self-diffusion coefficient
\be
D=\frac{l^2}{2\tau}\,. \label{D_CTRW}
\ee

The parameters of the model can be further constrained by imposing that $G_s$ be consistent with the van Hove function of a liquid.
To this end, we recall that, in a liquid, the msd and the VAF are related by the relation (\ref{general_relation_msd_vaf}), and that the self-diffusion coefficient of a liquid is related to the VAF via the Kubo relation $D=\int_0^\infty{Z(t)dt}$.
Using Eqs.(\ref{msd_model}) and (\ref{ZvandWv}) in (\ref{general_relation_msd_vaf}), we are led to interpret
\be
Z(t)=4\frac{d^2 W_v^2(t)}{dt^2}e^{-t/\tau}=Z_v(t)e^{-t/\tau}\,. \label{Zt}
\ee
as the VAF consistent with our model.
This is nothing but the celebrated VAF model of Zwanzig mentioned earlier, which allows us to connect to the numerous works that this model initiated.
Now, by enforcing the Kubo relation between Eqs.(\ref{Zt}) and (\ref{D_CTRW}), we find the constraint
\be 
l^2=\frac{2k_BT}{m}\tau^2\int_0^\infty{d\omega \frac{\rho(\omega)}{1+(\tau\omega)^2}} \label{l2Kubo}
\ee
between the three original parameters of the model.

In summary, our model for the first exit time properties works as follows.
Given an input density of state $\rho(\omega)$ and an input jump time $\tau$, the probability density $G_s({\bf r},t;{\bf r}_0)$ of finding an atom with position ${\bf r}$ at time $t$ if it was initially at ${\bf r}_0$ is given by Eq.(\ref{G_s}) (or (\ref{Gs_ks_model})), in which the average size of jumps is given by Eq.(\ref{l2Kubo}).
The resulting $G_s$ is then substituted into the integral equation (\ref{markovian_eq_for_f_integral}), the solution of which gives the desired probability of first exit times.

\subsection{Application} \label{section_4_1_applications}

In the following, we find it convenient to use the cage size $r_v$, Eq.(\ref{r_v}), as the unit of length, and the inverse of the root mean square frequency $\nu_0=\omega_0/2\pi$ with $\omega_0^2=\langle\omega^2\rangle$ as the unit of time; the model then depends on the dimensonless quantities $\tilde\rho=\nu_0\rho$ and $\tilde\tau=\nu_0\tau$.

In order for the model calculations to be as realistic as possible, we present results obtained with the frequency distributions $\rho(\omega)$ calculated by Chisolm et al. for liquid Na  (see Fig.~1 in Ref.~\cite{ChisolmClementsWallace2001}) and reproduced in Fig.~\ref{Fig:DofR_random_walk_Na} (symbols).
The spectrum was calculated with molecular dynamics simulations by carefully quenching the liquid into several stable random valleys of the potential energy surface, by calculating the normal mode frequency spectrum at the bottom of each valley and by averaging over the valleys \cite{WallaceClements1999,ChisolmClementsWallace2001}.

Figure~\ref{Fig:Dr_Na_rabani} (upper panel) shows the logarithm derivative $D(r)=\frac{d\ln \tau}{d\ln r}$ as a function of $r/r_v$ for several increasing values of the jump time $\nu_0\tau$ in the range $[.,.]$, i.e. for decreasing temperature $T$.
We first note that, overall, the model reproduces the typical behavior of $D(r)$ found in the simulations (see Sec.~\ref{section_2}) and exhibits the emergence of the peak in $D(r)$ of increasing height as $\nu_0\tau$ increases.
The signature peak value $D(r_*)=2.1$ of the freezing transition discussed in Sec.~\ref{section_2} is obtained for $\nu_0\tau=$ (highlighted in red in Figure~\ref{Fig:Dr_Na_rabani}) and is reached at a distance $r_* \simeq r_v-1.1 r_v$, i.e.  near the size of the cage.
Therefore, the model suggests that the dynamical signature of the liquid transition corresponds to conditions under which the jump time between valleys is equal to the typical period of oscillation of atoms around their instantaneous equilibrium position.
It suggests that the location of the freezing transition is concomitant with a crossover in the degree of localization of liquid particles.
Right above (below) the freezing temperature, the liquid configurations transit between the valley of the potential energy surface at a rate greater (smaller) than the average period of vibration in a given valley.
The two time scales coincide at the transition.

Of course, the previous conclusions will apply to other liquids only if the power index $D(r)$ predicted by the model is insensitive to the input frequency spectrum $\rho(\omega)$, which is also the condition that the model be consistent with the universality observed in the simulations.
To our knowledge, only a small number of spectra have been reported in the literature.
In Ref.~\cite{RabaniGezelterBerne1997}, Rabani et al. presented several calculations of $\rho(\omega)$ for Lenard-Jones liquids at different densities.
The Lennard-Jones spectra of Rabani et al. (see Fig.~7 in Ref.\cite{RabaniGezelterBerne1997}) show strong resemblance to the Na spectrum.
They all show similar bump-like shapes and linear behavior at low frequency; they differ mostly at high frequency, e.g. in the speed at which the spectra vanish beyond a cutoff frequency.
This is illustrated in Fig.~\ref{Fig:DofR_random_walk_Na}.
The dashed blue line shows the one-parameter fitting formula
\be
\rho(\omega)=\frac{2\omega}{\omega_0^2}e^{-\omega^2/\omega_0^2}\,, \label{Rabani_et_al_model}
\ee
proposed by Rabani et al. to model their simulation data (see Fig.~7 in Ref.\cite{RabaniGezelterBerne1997}).
Unfortunately, while the formula does reproduce well their data, we find that Eq.(\ref{Rabani_et_al_model}) is unphysical at very small $\omega$ where it varies linearly with $\omega$ and causes the cage size $r_v$, Eq.(\ref{r_v}) to diverge logarithmically to infinity.
Close inspection of Fig.~7 in Ref.\cite{RabaniGezelterBerne1997} shows that, like the Na spectrum, $\rho(\omega)$ varies linearly but only beyond some finite frequency ($\sim 0.05\omega_0$ for the Na spectrum) and below which the spectrum is vanishingly small.
The spectrum obtained by slightly shifting Eq.(\ref{Rabani_et_al_model}) to reproduce the linear section of the Na spectrum is shown by the red line; it deviates mostly from the Na spectrum at high frequencies.
Figure~\ref{Fig:DofR_random_walk_Na} (bottom panel) shows the logarithm derivative $D(r)=\frac{d\ln \tau}{d\ln r}$ as a function of $r/r_v$ obtained with this model spectrum for the same values of $\nu_0\tau$ used in the upper panel.
We see that despite the differences in the spectra at high $\omega$, the $D(r)$'s in both cases are very similar and lead to the same conclusions as before.
Thus, the model suggests that the insensitivity of the power index $D(r)$ to the nature of particle interactions observed in the numerical simulations results from the insensitivity of frequency spectra $\rho(\omega)$ at low frequency.
Additional calculations of $\rho(\omega)$ are needed to support this conclusion.

\begin{figure*}[t]
\begin{center}
\includegraphics[scale=0.29]{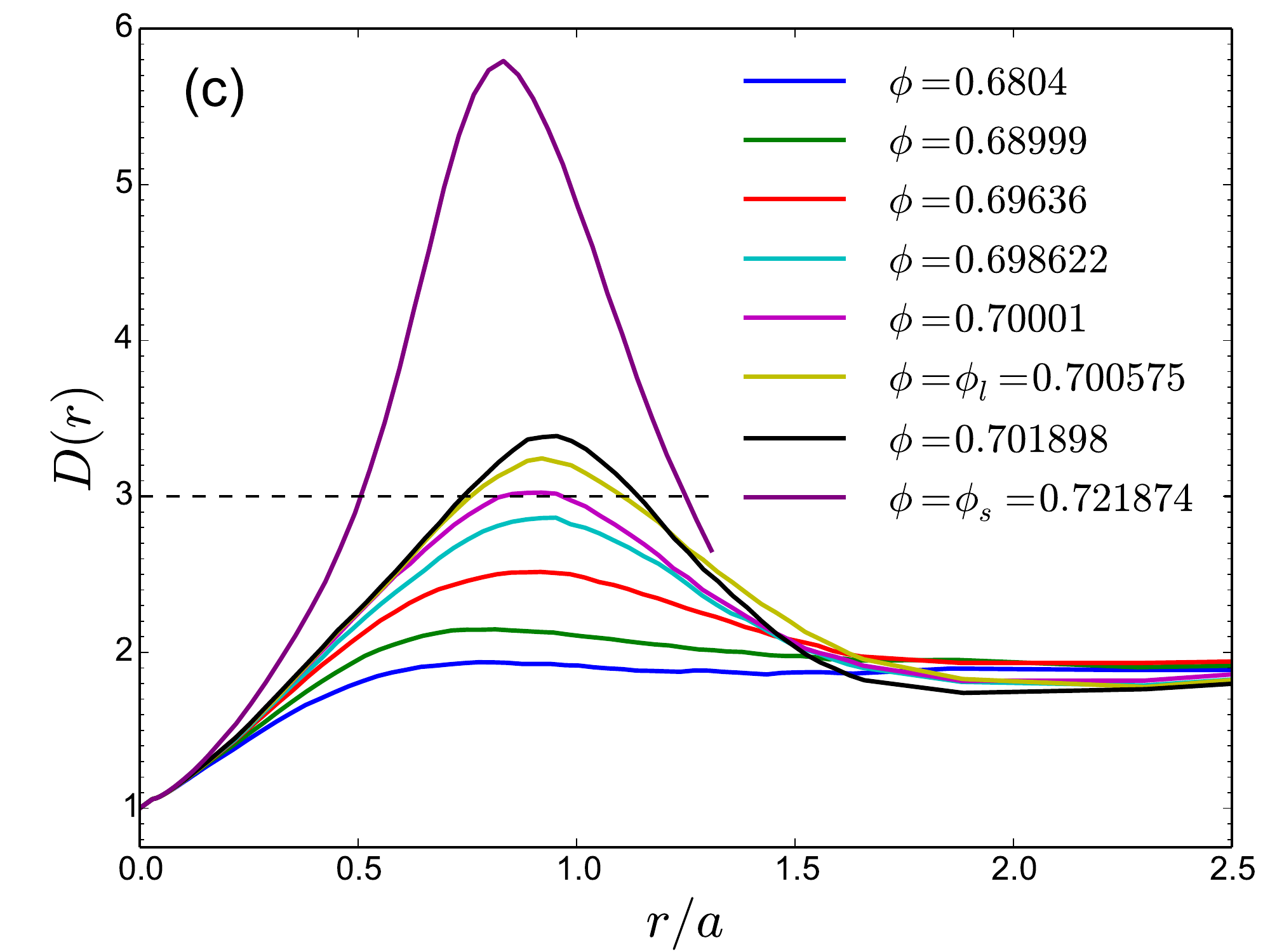}
\includegraphics[scale=0.29]{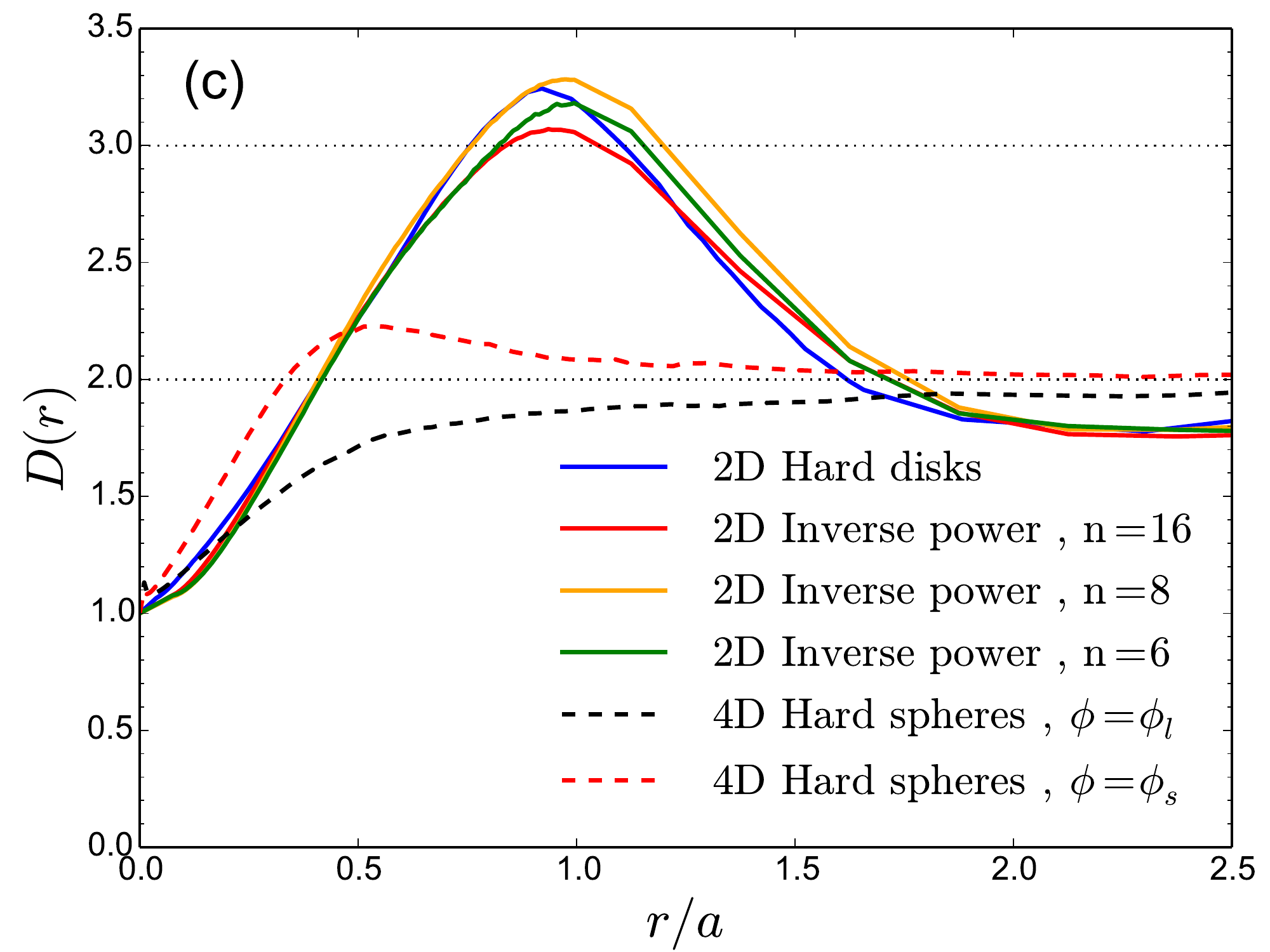}
\includegraphics[scale=0.29]{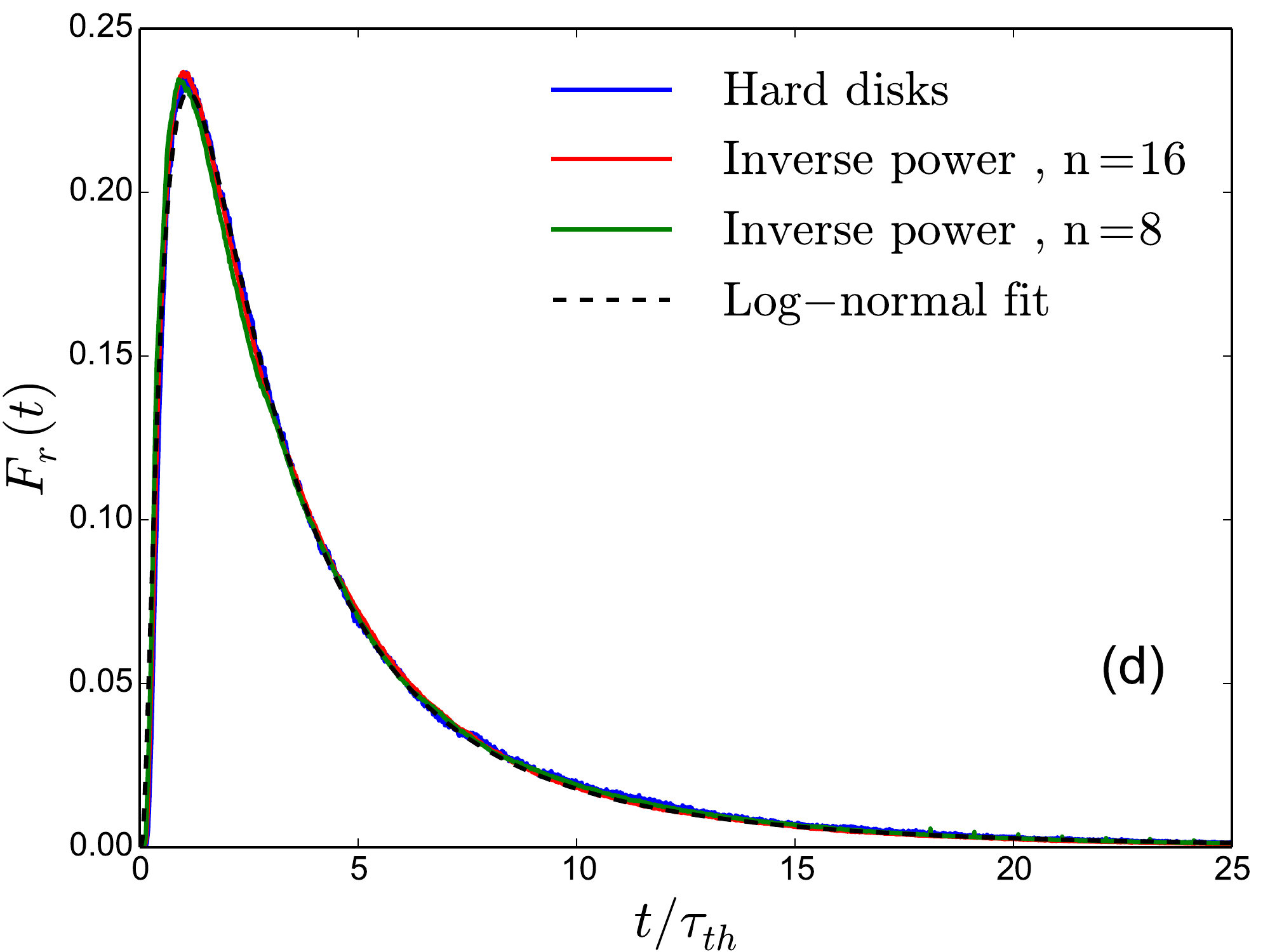}
\end{center}
\caption{(color online) (a) Logarithmic derivative $D(r)=d \ln{\cal{T}}\!(r)/d\ln(r)$  of the mean first exit time for several reference models of 3D liquids at their freezing point. The freezing points were determined by others from accurate free-energy calculations (see appendix~\ref{appendix_1}). (b) First-exit time probability distributions $f(r,t)$ for the 3D liquids of panel (a) measured at the location $r=r^*$ of the maximum of $D(r)$ (for clarity, only a subset of the cases in panel (a) are shown). The dashed line is a least-square fit to a log-normal distribution function. (c) Same as panel (a) for 2D and 4D systems. (d) Same as panel (b) for the 2D systems of panel (c).
\label{figure2}}
\end{figure*}

\section{Effect of dimensionality} \label{section_5}

In order to further investigate the relation between freezing point and particle localization, we study the effect of the dimension of space $d$. Dimensionality is indeed known to strongly affect nucleation in particular through its effect on the geometric packing, the geometric frustration and the thermal fluctuations. We first consider two-dimensional (2D) systems, starting with the hard-disk model (see appendix~\ref{appendix_1}). Its phase diagram \cite{17} consists of a fluid phase for packing fractions $\phi\le\phi_l=0.7006$ and a solid phase for $\phi\ge \phi_s=0.7218$, with $\phi=\pi\rho\sigma^2/4$. These are separated by a first-order liquid-hexatic transition at $\phi_{hex}=0.7171$ and a continuous hexatic-solid transition at $\phi_s$, where the hexatic phase is characterized by short-range positional and quasi-long-range orientational orders. Figure 1c shows $D(r)$ for $0.6804\le \phi\le\phi_s$. Remark that $D(r)$ is extremely sensitive to $\phi$; for instance, the packing fractions $\phi=0.70001$ (purple line) and $\phi=0.701898$ (black) are only within $0.08\%$ and $0.19\%$ of $\phi_l$ (yellow). For $\phi=\phi_l$ , $D(r)$ peaks at $\sim 3.2$ instead of $2.1$ and the new criterion is not applicable in two dimensions. Similarly, it is known that the 3D Hansen-Verlet and the Lindemann criteria do not carry over to two dimensions \cite{18,19,20}.

To investigate whether 2D freezing can nevertheless be characterized by a common localization threshold, Figure 2c shows $D(r)$ at the freezing conditions of four models of inverse power potentials with stiffness $n=\infty,16,8$ and $6$ (hard disks correspond to $n=\infty$). Their phase diagrams \cite{21} are similar to that of hard disks except for the location of the coexistence lines. Like with 3D systems, the strong similarity between the plots is remarkable. For $n=8$ and $6$, the peak height is within $< 2\%$ that of hard-disks. For $n=16$, the peak is $5\%$ lower, a larger disagreement that we ascribe to the above-mentioned high sensitivity on $\phi$ and the inaccuracy (estimated to \cite{21} $0.5\%$) in $\phi_l$. In addition, Figure 2d shows that the corresponding probability distributions $f(r,t)$  against $t/t_{th}$ at peak position $r^*$ of $D(r)$ are in remarkably close agreement (in 2D, $t_{th}=a/v_{th}$  with $a=1/\sqrt{\pi\rho}$). Thus, we again find a strong correlation between the location of the freezing transition of 2D fluids and the onset of a regime of localization of atomic motions characterized by $D(r^*)\simeq 3.2$, different from the 3D threshold. The reduction of the threshold with the space dimension d suggests that the relationship between freezing conditions and particle localization is specific to low dimensions $d=2,3$. To check this hypothesis, Figure 2c (dashed lines) shows $D(r)$ for a fluid of hard hyperspheres in $d=4$ dimensions at the fluid-solid coexistence values \cite{22} $\phi_l=0.288$ and $\phi_s=0.337$, with $\phi=\pi^2 \rho\sigma^4/32$. For $\phi=\phi_l$, $D(r)$ increases strictly monotonically between the inertial and diffusive regimes, while for $\phi=\phi_s$, $D(r)$ shows a small hump of height $2.2$. It is known that hyperspheres do not freeze easily \cite{22}. The barrier of crystal nucleation was shown to rapidly grow with $d$ as a result of the increased geometrical frustration between the fluid order and the crystal \cite{22}. The present work suggests that the difficulty to freeze is also related to the weak localization of particle motions that prevents the many-body interactions needed to form stable solid nuclei.

\section{Conclusion} \label{section_6}

While the principles of thermodynamics unambiguously predict the location of the liquid-solid transition by the strict conditions of equality of the pressure, temperature, and chemical potentials of both phases pure liquids can generally be super-cooled or over-compressed passed these conditions with no sign of abrupt changes in their properties around the transition.
The results of this work provide new insights into what changes at the atomic level as the freezing point is traversed and allows the system to discover the crystalline order.

We have considered the first passage time properties of atomic motions, with a special attention on the mean time ${\cal{T}}(r)$ for an atom to first reach a distance $r$ from its initial position and the associated probability distribution.
We have shown evidence from classical and quantum molecular dynamics simulations of a universal feature exhibited by ${\cal{T}}(r)$ in monatomic liquids that heralds the freezing transition.
For temperatures above freezing, the mean first passage time ${\cal{T}}(r)$ behaves as $r^{D(r)}$ with a power index $D(r)$ that monotonically increases from $D(r)\!=\!1$ at small $r$ (free-particle behavior) to $D(r)\!=\!2$ at large $r$ (diffusive behavior).
At freezing, and regardless the nature of interactions, $D(r)$ no longer varies monotonically between these two values but
exhibits a peak of height $D(r_*)\!=\!2.1$ at some distance $r_*$.
Presented numerical evidence includes data for several reference models of liquids spanning from the hard-sphere fluid to the one component plasma model, as well as data for real liquids metals obtained with quantum molecular dynamics simulations.
We have shown that the precursory feature induces a new method for determining the liquid-solid coexistence curves of real materials from atomistic simulations.
We have successfully illustrated the method on the calculation of the solid-liquid coexistence curves of liquid Aluminum and liquid Titanium.
Unlike other methods, the methods views crystallization from the liquid side and does not require knowing the crystalline structure of the solid phase.
This is evocative of the result of Alexander and McTague \cite{AlexanderMcTague1978} based on general symmetry considerations of the liquid-solid transition that the bcc crystal should be favored near the melting line.

Then, in order to help understand and characterize the physics underlying our finding, we have developed a model of the first-passage properties of atomic motions in liquids based on the potential energy landscape theory for liquids.
The model combines an accurate description of the localized oscillations of an atom about an equilibrium position together with a continuous time random walk to account for the occasional jumps that occur between equilibrium positions.
The model depends on only two physical parameters, namely the average frequency spectrum $\rho(\omega)$ of normal modes in a local minimum and on the average time $\tau$ between jumps.
We have applied the model to realistic frequency spectra and have shown that it reproduces the variations of the first passage time properties observed in the computer simulations.
The model implies that the freezing point is concomitant with a change in the degree of localization of atoms.
At the freezing point, the average time $\tau$ separating two transits is equal to the average period of oscillation $\tau_o$ of an atom about an equilibrium position; below (above) the freezing point, $\tau$ is larger (smaller) then $\tau_o$, i.e. atoms remain localized for times longer (shorter) than the typical period of oscillations in a local potential energy valley.
The longer localization in the valleys of the potential energy surface is a necessary condition for atoms to interact constructively and find the route to a local crystalline order.

Although consistent with the hard-sphere paradigm of liquids \cite{Dyre2016}, we believe that the present findings are remarkable in view of the non-univeral character of the freezing transition \cite{3}.
The properties of freezing such as its location, the changes of thermodynamic variables and the crystalline structure selected, depend indeed sensitively on the nature of the intermolecular forces.
In addition, unlike critical phenomena, one cannot restrict attention to long-wavelength phenomena since the formation of localized solid nuclei likely depends on the small scales given by the range of intermolecular interactions \cite{Binder1987}.
We hope that this work will stimulate further research to elucidate the mechanisms that govern the liquid-solid phase transition on the microscopic level and their potential implications on the conventional nucleation theory.

\acknowledgments
The author thanks Dr. Didier Saumon for useful discussions and for his encouragements.
This work was performed under the auspices of the U.S. Department of Energy under Contract No. 89233218CNA000001.

\appendix

\section{Models of simple liquids used in this work.} \label{appendix_1}

We have considered homogeneous
systems of identical particles interacting through a pair-potential
$v(r)$ , where r is the interparticle distance. In the hard-sphere
model\cite{25}, impenetrable spheres of diameter $\sigma$ mutually interact via the
repulsive potential
\be
v(r)=\left\{\begin{array}{l}
0\quad\text{for }r>\sigma\\
\infty\quad\text{for }r\leq\sigma
\end{array}
\right. .
\ee
Its equilibrium properties are fully characterized by the packing
fraction $\phi=\Omega_0/\Omega$, ratio of the volume of a particle $\Omega_0$ and the
volume per particle $\Omega=1/\rho$. We find $\phi=\pi\rho\sigma^3/6$ for 3D spheres,
$\phi=\pi\rho\sigma^2/4$ for 2D disks and $\phi=\pi^2 \rho\sigma^4/32$ for 4D
hyperspheres\cite{22}. In the inverse power or soft sphere
model\cite{26} 
\be
v(r)=\epsilon \left(\frac{r}{\sigma}\right)^n
\ee
where $\epsilon$  is an energy scale and $\sigma$ is the effective parameter of
a particle. The exponent $n$ controls the stiffness and range of the
repulsion, from the short-ranged hard-sphere interaction ($n=\infty$) to
the long-range Coulomb interaction ($n=1$). Due to scaling properties,
the thermodynamic properties are fully characterized by the
dimensionless parameter $\gamma_n=\rho\sigma^d (k_B T/\epsilon)^(-d/n)$, where d is the
space dimension. For fixed $\epsilon,\gamma_n$ can be replaced by the packing
fraction defined above. The Lennard-Jones model\cite{25} is often used
to model fluids made of neutral atoms or small molecules. The
potential
\be
v(r)=4\epsilon[(r/\sigma)^{12}-(r/\sigma)^6 ]  
\ee
consists of a short-range repulsive term and longer ranged, attractive
part. Here $\sigma$ is the atomic diameter and $\epsilon$  the depth of the
attractive well. The one-component plasma model\cite{27} is a system of
charged particles of electric charge $q$ immersed in a homogeneous
neutralizing background and interacting through the Coulomb potential
\be
v(r)=\frac{q^2}{4\pi\epsilon_0 r}
\ee
Its thermodynamic properties are fully characterized by the Coulomb
coupling parameter $\Gamma=(q^2/(4\pi\epsilon_0 a))/k_B T$, where $a=(3/4\pi\rho)^{1/3}$
is the interparticle distance. The one-component plasma is often used
to model the ions in dense, strongly coupled plasmas as those found in
the core of astrophysical objects. In the Yukawa model\cite{28}, the bare
Coulomb interaction is exponentially screened, 
\be
v(r )=\frac{q^2}{4\pi\epsilon_0 r} e^{-\kappa r/a}
\ee
where the inverse screening length (in units of $1/a$) $\kappa$ describes the screening effect of plasma electrons on the bare ion-ion Coulomb interactions.
The effect of $v(r)$ on the single particle dynamics is illustrated in figure 1a. For instance, for hard spheres, $Z(t)$ rapidly vanishes after the first rebound against the initial cage, while for the Lennard-Jones interaction, $Z(t)$ oscillates with larger negative correlations than for hard-spheres. For the Coulomb ($\kappa=0$) one-component plasma, unlike other models, the lowest minimum of $Z(t)$ is attained by its second minimum. This is because, in addition to the oscillatory motions in the cages, particles also couple to the collective, high-frequency (plasma) charge oscillations\cite{29}; this effect disappears with increasing $\kappa$  as the plasma oscillations are replaced by low-frequency sound waves. Very similar temporal variations (not shown here) are found in both the stable and metastable liquids in neighbourhood of the transition with no clear signature of a change of behaviour at the transition. 

The freezing and melting conditions used in this work
were determined by others from accurate free-energy calculations. For
the hard and soft sphere models, we used
\begin{center}
\begin{tabular}{c|c||c|c|c||c}
$d$ & $n$ & $\phi_l$ & $\phi_{hex}$ & $\phi_s$ & Reference\\\hline
2 & $\infty$ & 0.7006 & 0.7171 & 0.7218 & Table I in [21] \\
2& $16$       & 0.7359 & 0.7453 & 0.7540 & Table I in [21]  \\
2 & $6$        & 1.1282 & 1.1836 & 1.1906 & Table I in [21] \\\hline
3 & $\infty$ & 0.494 & X & 0.545 & Table I in [22]  \\\hline 
4 & $\infty$ & 0.288 & X & 0.337 & Table I in [22] 
\end{tabular}
\end{center} 
For the one-component plasma and Yukawa models, we used
\begin{center}
\begin{tabular}{c||c|c|c|c|c}
$\kappa$ & 0 & 2 & 3 & 4 & 4.6 \\\hline
$\Gamma$ & 175  & 440.1 & 1185 & 3837 & 8609
\end{tabular}
\end{center}
given in Table X of Ref. [28]. For the Lennard-Jones systems, we used
equation 3 of Ref. [30].

\section{Molecular dynamics (MD) calculations}

\subsection{Classical MD}

All the simulations were performed with computer codes written entirely by the author. Briefly, in all cases, $N$ particles are evolved in a cubic box of volume V, and periodic conditions are imposed on all boundaries.
The simulations of hard sphere in 2D, 3D and 4D were performed with a standard event driven algorithm that evolves the system on a collision-by-collision basis\cite{31}, computing the collision dynamics and then searching for the next collision.
The initial random packings are generated using the algorithm proposed by Julien et al.\cite{32}.
For the other (continuous) potentials, the particle dynamics is obtained by solving Newton's
equations of
motion with the Verlet integrator.
For the one-component plasma, the forces are calculated using the Ewald summation technique.
For numerical efficiency, the latter is calculated with a parallel implementation of the particle-particle-particle-mesh method that simultaneously provides high resolution for individual encounters combined with rapid, mesh-based, long range force calculations\cite{31}.
For the short-ranged potentials, the force calculations are performed using standard neighbouring list techniques. 
The simulation requirements to calculate the first-exit time
properties are standard. A typical simulation consists of an
equilibration phase of length $t_{eq} = N_{eq}\delta t$ (only for
continuous potentials) followed by the main MD run of length $t_{MD} =
N_{MD} \delta t$ for a total of $N_{eq} + N_{MD}$ time steps. During
the equilibration phase, velocity scaling is used to maintain the
desired temperature. Velocity scaling is turned off after the
equilibration phase. For the continuous potentials, the time step is
chosen to ensure good energy conservation (one part in a million),
typically $\delta t=0.01t_{th}$. For the stiff inverse power potential
($n\ge 6$) we used $\delta t=0.001t_{th}$ to ensure a good description
of close collisions. We used $N=1024$ particles for the 3D and 4D
simulations and $N=625$ for the 2D simulations. No significant change
was found when using more particles. All particles were used for the
calculation of first passage properties. The simulation length must be
long enough to ensure that the vast majority of particles travel far
away from their initial positions (several times the maximum distance
r used to calculate ${\cal{T}}(r)$. This is to improve the statistics of
exit times discussed below.

\subsection{Quantum MD}

The QMD simulations of liquid Al and liquids Ti were performed with the open-source Quantum-Espresso program \cite{Giannozzi2009}
with standard numerical parameters appropriate for these elements (e.g., Ref.~\cite{Bouchet2009,Stutzmann2015}).
A detailed study similar to that presented in \cite{Bouchet2009} for the $Z-method$ and the coexistence method on the influence of numerical parameters on the distribution of first passage times $f(r,t)$ and on the power index $D(r)$ is beyond the scope of this work.
Brifely, the electronic structure is obtained by solving the finite temperature Kohn-Sham equation in a plane-wave basis at the $\Gamma$-point only and with the exchange-correlation potential of Perdew, Burke, and Ernzerhof.
A projector augmented-wave (PAW) pseudopotential was used to describe the electron-ion interactions.
Simulations are performed in the NVT ensemble with a Nose-Hoover thrermostat.
In all cases the simulations included 64 atoms in the unit cell, with time steps of $1$ fs, and over a time duration of $10$ fs.

\section{Calculation of first exit time properties}

Consider the spatial
trajectory $\vec{R}_i(t)$ of a given particle i as a function of time $t$,
which starts at $\vec{R}_i(t_0)$ with velocity $\vec{V}_i(t_0 )$ at initial time $t_0$. The exit time from a spherical domain of radius r is defined as the first time the particle reaches any point at a distance r from its starting point $\vec{R}_i(t_0)$, i.e.
\ben
\lefteqn{\tau\left(r ; i,\vec{R}_i(t _0),\vec{V}_i(t_0)\right)}&&\\
&=&{\rm inf}\left\{t\geq t_0\,:\, |\vec{R}_i(t)-\vec{R}_i(t_0)|^2>r\right\} .
\een
The first exit time becomes a random variable when considered over the
set of all particles $1\leq i\leq N$ and a thermal ensemble of initial
conditions. In practice, in the molecular dynamic simulations, $N$
particle trajectories $\vec{R}_i(t_n )$ are calculated and stored at
discrete time steps $t_n=n\times\delta t$ with $0\leq n\leq
N_{MD}$. To greatly improve the statistics, assuming ergodicity, each
time step $t_n$ can be regarded as the initial time of N new
trajectories with initial conditions $\{\vec{R}_i(t_n ),\vec{V}_i(t_n
)\}_{i=1,…,N}$. For each particle $i$ and initial time $t_n$, the first
exit time from a distance $r$ from the initial position $\vec{R}_i(t_n
)$ is then given by $m_r^*(i,n)\times\delta t$ with
\ben
\lefteqn{m_r^* (i,n)}&&\\
&&={\rm inf}\left\{1\leq m\leq N_{MD}+1:|\vec{R}_i(t_{n+m})-\vec{R}_i(t_n )|^2>r \right\}\,.
\een
When this is calculated for all particles $1\leq i\leq N$, and for all initial
times $n \delta t$ with $1\leq n\leq N_{MD}$, we obtain the probability distribution
$f(r,t)$ of first exit time by storing the first exist times $m_r^* (i,n)$ in an histogram and then by evaluating
\ben
\lefteqn{f(r,k \delta t)\delta t}&&\\
&&=\frac{\sum_{i=1}^N{\sum_{n=0}^{N_{steps}}{ \theta\left(N_{MD}-m_r^*(n,i)\right)\delta\left(m_r^*(n,i),k\right)}}}{\sum_{i=1}^N{\sum_{n=0}^{N_{steps}}{\theta\left(N_{MD}-m_r^* (n,i)\right)}}}\,,
\een
where $\delta(a,b)$ is the Kronecker delta, and $\theta(x)$ is the
Heaviside function. $f(r,k \delta t)\delta t$ corresponds to the probability
that a particle of the liquid reaches the distance r from its current
position between times $t=k \delta t$ and $t+\delta t$. The denominator provides the normalization of the probability density.


\begin{references}
\bibitem{1}  D.W. Oxtoby, {\it Homogeneous nucleation: theory and experiment}, J. Phys.: Condens. Matter {\bf 5}, 7627 (1992). 
\bibitem{7} P. Papon, J. Leblond and P.H.E. Meijer, {\it The Physics of Phase Transitions, Concepts and Applications, 2nd Edition} (Springer, 2006). Chapter 3.
\bibitem{3} D.W. Oxtoby, {\it New perspectives on freezing and melting}, Nature {\bf 347}, 725 (1990). 
\bibitem{5} P.G. Debenedetti, {\it Metastable Liquids Concepts and Principles} (Princeton Univ Press, Princeton), pp 146-199. 
\bibitem{4} V.A. Martinez, E. Zacccarelli, E. Sanz, C. Valeriani and W. van Megen, {\it Exposing a dynamical signature of the freezing transition through the sound propagation gap}, Nat. Commun. {\bf 5}:5503 doi: 10.1038/ncomms6503 (2014). 
\bibitem{6} T.M. Truskett, S. Torquado, S. Sastry, P.G. Debenedetti and F.H. Stillinger, {\it Structural precursor to freezing in the hard-disk and hard-sphere systems}, Phys. Rev. E {\bf 58}, 3083 (1998).
\bibitem{Giaquinta1992} P.V. Giaquinta, G. Giunta, and S. Prestipino Giarritta, {\it Entropy and the freezing of simple liquids}, {\it Phys. Rev. A} {\bf 45}, R6966 (1992).
\bibitem{Lowenetal1993} H. L{\"o}wen, T. Palberg and R. Simon, {\it Dynamical Criterion for Freezing of Colloidal Liquids}, Phys. Rev. Lett. {\bf 70}, 1557 (1993). 
\bibitem{9} J.P. Hansen and L. Verlet, {\it Phase Transitions of the Lennard-Jones System}, Phys. Rev. {\bf 184}, 151 (1969).
\bibitem{10} J.L. Barrat and J.P. Hansen, {\it Basic Concepts for Simple and Complex Liquids} (Cambridge University Press, 2003). Chap. 4.6.
\bibitem{11} U. Balucani and M. Zoppi, {\it Dynamics of the Liquid State} (Oxford Science Press, 1994), Sec. 1.4.2.
\bibitem{11ocp} J. Daligault, {\it Liquid-state properties of a one-component plasma}, Phys. Rev. Lett. {\bf 96}, 065003 (2006).
\bibitem{25} Hansen, J.P. \& McDonald, I.R. {\it Theory of Simple Liquids with Applications to Soft Matter} (Academic Press, Fourth Edition, 2013).
\bibitem{13} P. Allegrini, J.F. Douglas and S.C. Glotzer, {\it Dynamic entropy as a measure of caging and persistent particle motion in supercooled liquids}, Phys. Rev. E {\bf 60}, 5714 (1999).
\bibitem{12} P.L. Krapivsky, S. Redner and E.A. Ben-Naim, {\it Kinetic View of Statistical Physics} (Cambridge University Press, 2010). Chap. 2.6.
\bibitem{15} J.J. Gilvarry, {\it The Lindemann and Gr{\"u}neisen Law}, Phys. Rev. {\bf 102}, 308 (1956).
\bibitem{16} See Table I. in \cite{Lowenetal70}.
  \bibitem{Bouchet2009} J. Bouchet, F. Bottin, G. Jomard, and G. Z{\'e}rah, {\it Melting curve of aluminum up to 300 GPa obtained through ab-initio molecular dynamics simulations},  {\it Phys. Rev. B} {\bf 80}, 094102 (2009).
  \bibitem{Stutzmann2015} V. Stutzmann, A. Dewaele, J. Bouchet, F. Bottin, and M. Mezouar, {\it High-pressure melting curve of titanium}, {\it Phys. Rev. B} {\bf 92}, 224110 (2015).
\bibitem{Boehler1997} R. Boehler and M. Ross, {\it Melting curve of aluminum in a diamond cell to 0.8 Mbar: implications for iron}, {\it Earth and Planetary Science Letters} {\bf 153}, 223 (1997).    
\bibitem{PolyaninManzhirov_book} A.D. Polyanin and A.V. Manzhirov, {\it Handbook of integral equations} (CPC Press, 1998).
\bibitem{details_on_diffusion_example} One finds $\hat g(r_*,s)=\frac{(r_*\sqrt{D/s}+1)}{s}e^{-r_*\sqrt{D/s}}$ and $\hat K(r_*,s)=\frac{{\rm sh}\left(r_*\sqrt{s/D}\right)} {r_*\sqrt{s/D}}\hat g(r_*,s)$.
  \bibitem{Klein1952} G. Klein, {\it Mean First-Passage Times of Brownian Motion and Related Problems}, {\it Proc. of the Roy. Soc. of London. Series A} {\bf 211}, 431 (1952).
  \bibitem{BorodinSalminenbook} A.B. Borodin and P. Salminen, {\it Handbook of Brownian Motions - Facts and Formulae} (Birkh{\"a}user Verlag, 1996).
  \bibitem{Rahman1964} A. Rahman, {\it Phys. Rev.} {\bf 136}, A405 (1964).
  \bibitem{LevesqueVerlet1970} D. Levesque and L. Verlet, {\it Phys. Rev. A} {\bf 2}, 2514 (1970).
    \bibitem{NijboerRahman1966} B.R.A. Nijboer and A. Rahman, {\it Physica} {\bf 32}, 415 (1966).
  \bibitem{Frenkelbook} J. Frenkel, {\it Kinetic Theory of Liquids} (Oxford University Press, 1946) [see Chapter III].
  \bibitem{StillingerWeber1984} F.H. Stillinger and T.A. Weber, {\it Packing structures and transitions in Liquids and Solids}, {\it Science} {\bf 225}, 983 (1984).
    \bibitem{Stillingerbook} F.H. Stillinger, {\it Energy Landscapes, Inherent Structures, and Condensed-Matter Phenomena} (Princeton University Press, 2015).
  \bibitem{Wallacebook} D.C. Wallace, {\it Statistical Physics of Crystals and Liquids} (World Scientific, Singapore, 2003).
  \bibitem{Sears1965} V.F. Sears, {\it The itinerant oscillator model of liquids}, {\it Proc. Phys. Soc. (London)} {\bf 86}, 953 (1965).
  \bibitem{Damle1968} P.S. Damle, A. Sj{\"o}lander, and K.S. Singwi, {\it Itinerant-Qscillator Model of Liquids}, {\it Phys. Rev.} {\bf 165}, 277 (1968).
    \bibitem{Zwanzig1983} R. Zwanzig, {\it J. Chem. Phys.} {\bf 79}, 4507 (1983). 
\bibitem{Keyes1997} T. Keyes, {\it Review on "Instantaneous Normal Mode Approach to Liquid State Dynamics}, {\it J. Phys. Chem.} {\bf 101}, 2921 (1997).
\bibitem{RabaniGezelterBerne1997} E. Rabani, J.D. Gezelter and B.J. Berne, {\it Calculating the hopping rate for self-diffusion on rough potential energy surfaces: Cage correlations}, {\it J. Chem. Phys.} {\bf 107}, 6867 (1997).
\bibitem{Wallace1997} D.C. Wallace, {\it Liquid dynamics theory of the velocity autocorrelation function and self-diffusion}, {\it Phys. Rev. E} {\bf 58}, 538 (1998).
\bibitem{ChisolmClementsWallace2001} E.D. Chisolm, B.E. Clements, and D.C. Wallace, {\it Mean-atom-trajectory model for the velocity autocorrelation function of monatomic liquids}, {\it Phys. Rev. E} {\bf 63}, 031204 (2001).
\bibitem{WeissRubin1983} G.H. Weiss and R.J. Rubin, {\it Random walks: Theory and selected applications}, {\it Adv. Chem. Phys.} {\bf 52}, 363 (1983).
\bibitem{note_on_fJ_phi} Note that many of the formulas below remain valid regardless of the functional form chosen for $f_J$ and $\phi$ and other forms could possibly be used.
  The choice of the exponential distribution not only gives a satisfactory physical model but also has the advantage to greatly simplifies the mathematical expressions and the numerical implementation.
\bibitem{Chaudhuri2007} P. Chaudhuri, L. Berthier and W. Kob, {\it Universal Nature of Particle Displacements close to Glass and Jamming Transitions}, {\it Phys. Rev. Lett.} {\bf 99}, 060604 (2007).  See Eq.(2) with $\tau_1=\tau_2$.
\bibitem{WallaceClements1999} D.C. Wallace and B.E. Clements, {\it Nature of the many-particle potential in the monatomic liquid state: Energetics, kinetics, and stability}, {\it Phys. Rev. E} {\bf 59}, 2942 (1999).
\bibitem{17} E.P. Bernard and W. Krauth, {\it Two-Step Melting in Two Dimensions: First-Order Liquid-Hexatic Transition}, Phys. Rev. Lett. {\bf 107}, 155704 (2011).
\bibitem{18} J.M. Caillol, D. Levesque, J.J. Weis and J.P. Hansen, {\it A monte Carlo Study of the Classical Two-Dimensional One-Component Plasma}, J. Stat. Phys. {\bf 28}, 325 (1982).
\bibitem{19} Z. Wang, A.M. Alsayed, A.G. Yodh and Y. Han, {\it Two-dimensional freezing criteria for crystallizing colloidal monolayers}, J. Chem. Phys. {\bf 132}, 154501 (2010), and references therein.
\bibitem{20} K.J. Strandburg, {\it Two-dimensional melting}, Rev. Mod. Phys. {\bf 60}, 161 (1988).
\bibitem{21} S.D. Kapfer and W. Krauth, {\it Two-Dimensional Melting: From Liquid-Hexatic Coexistence to Continuous Transitions}, Phys. Rev. Lett. {\bf 114}, 035702 (2015).
\bibitem{22} J.A. Van Meel, B. Chardonneau, A. Fortini and P. Charbonneau, {\it Hard sphere crystallization gets rarer with increasing dimension}, Phys. Rev. E {\bf 80}, 061110 (2009).
\bibitem{26} Hoover, W.G., Gray, S.G. \& Johnson, K.W. {\it Thermodynamics properties of the fluid and solid phases for inverse power potentials} J. Chem. Phys. 55, 1128 (1971)
  \bibitem{27} Baus, M. \& Hansen, J.P. {\it Statistical Mechanics of Simple Coulomb Systems} Phys. Rep. 59, 1 (1980). 
\bibitem{28} Hamaguchi, S., Farouki, R.T. \& Dubin, D.H.E. {\it Triple point of Yukawa systems} Phys. Rev. E 56, 4671 (1997).
\bibitem{29} Daligault, J. {\it Liquid-state properties of a one-component plasma} Phys. Rev. Lett. 96, 065003 (2006).
\bibitem{30} Khrapak, S.A., Chaudhuri, M. \& Morfill, G.E. {\it
    Liquid-solid phase transition in the Lennard-Jones systems} Phys. Rev B 82, 052101 (2010)
\bibitem{31} Allen, M.P. \& Tildesley, D. {\it Computer Simulation of Liquids, Second Edition} (Oxford University Press, 2017).
\bibitem{32} Julien, R., Jund, P., Caprion, D. \& Quitmann, D. {\it Computer investigation of long-range correlations and local order in random packings of spheres} Phys. Rev. E 54, 6035 (1996).
\bibitem{Giannozzi2009} P. Giannozzi, S. Baroni, N. Bonini, M. Calandra, R. Car,
  C. Cavazzoni, D. Ceresoli, G.L. Chiarotti, M. Cococcioni, I. Dabo, A. Dal Corso, S. Fabris, G. Fratesi, S. de Giron- coli, R. Gebauer, U. Gerstmann, C. Gougoussis, A. Kokalj, M. Lazzeri, L. Martin-Samos, N. Marzari, F. Mauri, R. Mazzarello, S. Paolini, A. Pasquarello, L. Paulatto, C. Sbraccia, S. Scandolo, G. Sclauzero, A.P. Seitsonen, A. Smogunov, P. Umari, R.M. Wentzcovitch, {\it QUANTUM ESPRESSO: a modular and open-source software project for quantum simulations of materials}, {\it J. Phys.: Condens. Matter} {\bf 21}, 395502 (2009).
\bibitem{AlexanderMcTague1978} S. Alexander and J. McTague, {\it Should All Crystals Be bcc? Landau Theory of Solidification and Crystal Nucleation}, {\it Phys. Rev. Lett.} {\bf 41}, 702 (1978).
\bibitem{Dyre2016} J.C. Dyre, {\it Simple liquids’ quasiuniversality and the hard-sphere paradigm}, {\it J. Phys.: Condens. Matter} {\bf 28}, 323001 (2016).
\bibitem{Binder1987} K. Binder, {\it Theory of first-order phase transitions}, {\it Rep. Prog. Phys.} {\bf 50}, 783 (1987).
\end{references}
\end{document}